\documentclass[a4paper,11pt]{article}
\usepackage{jcappub} 
\usepackage{lineno}
\usepackage{color}
\usepackage{xspace}
\usepackage{manyfoot}
\usepackage{url}
\usepackage{bm}
\usepackage{makecell}
\usepackage{soul}
\usepackage{multirow}

\usepackage{multirow}
\usepackage{subcaption}

\usepackage{array}
\makeatletter
\newcommand{\thickhline}{%
    \noalign {\ifnum 0=`}\fi \hrule height 1pt
    \futurelet \reserved@a \@xhline
}
\newcolumntype{"}{@{\hskip\tabcolsep\vrule width 1pt\hskip\tabcolsep}}
\makeatother

\definecolor{ForestGreen}{rgb}{0.13, 0.55, 0.13}
\definecolor{airforceblue}{rgb}{0.36, 0.54, 0.66}
\definecolor{orange}{rgb}{1.0, 0.5, 0.0}
\definecolor{amethyst}{rgb}{0.6, 0.4, 0.8}
\definecolor{awesome}{rgb}{1.0, 0.13, 0.32}
\definecolor{chromeyellow}{rgb}{1.0, 0.65, 0.0}

\newcommand{\quijote}{\textsc{Quijote}\xspace}
\newcommand{\quijotepng}{\textsc{Quijote-PNG}\xspace}

\newcommand{\cnn}{\textsc{CNN}\xspace}
\newcommand{\mlp}{\textsc{MLP}\xspace}
\newcommand{\hybrid}{\textsc{Hybrid}\xspace}
\newcommand{\xgboost}{\texttt{XGBoost}\xspace}
\newcommand{\gbt}{\textsc{GBT}\xspace}

\newcommand{\fnlloc}{f_{\rm NL}^{\rm loc}}
\newcommand{\fnleq}{f_{\rm NL}^{\rm equi}}

\title{Cosmology with persistent homology: parameter inference via machine learning}

\footnotetext[1]{Equal contribution.}

\author[a]{Juan Calles,$^{*,}$}
\author[b]{Jacky H. T. Yip,$^{*,}$}
\author[c,d]{Gabriella Contardo,}
\author[e]{Jorge Nore\~na,}
\author[b]{Adam Rouhiainen,}
\author[b]{and Gary Shiu}
\affiliation[a]{Instituto de F\'isica y Astronom\'ia, Universidad de Valpara\'iso,\\Avda. Gran Bretaña 1111, Valpara\'iso, Chile}
\affiliation[b]{Department of Physics, University of Wisconsin-Madison,\\Madison, WI 53706, USA}
\affiliation[c]{Center for Astrophysics and Cosmology, University of Nova Gorica,\\Ajdovščina I-5270, Slovenia}
\affiliation[d]{Theoretical and Scientific Data Science, Scuola Internazionale Superiore di Studi Avanzati,\\Trieste 34136, Italy}
\affiliation[e]{Instituto de F\'isica, Pontificia Universidad Cat\'olica de Valpara\'iso, \\Casilla 4950, Valpara\'iso, Chile}

\abstract{
Building upon previous work \cite{Yip:2023vud}, we investigate the constraining power of persistent homology on cosmological parameters and primordial non-Gaussianity in a likelihood-free inference pipeline utilizing machine learning. We evaluate the ability of Persistence Images (PIs) to infer parameters, comparing them to the combined Power Spectrum and Bispectrum (PS/BS). We also compare two classes of models: neural-based and tree-based. PIs consistently lead to better predictions compared to the combined PS/BS for parameters that can be constrained, i.e., for $\{\Omega_{\rm m}, \sigma_8, n_{\rm s}, f_{\rm NL}^{\rm loc}\}$. PIs perform particularly well for $f_{\rm NL}^{\rm loc}$, highlighting the potential of persistent homology for constraining primordial non-Gaussianity. Our results indicate that combining PIs with PS/BS provides only marginal gains, indicating that the PS/BS contains little additional or complementary information to the PIs. Finally, we provide a visualization of the most important topological features for $f_{\rm NL}^{\rm loc}$ and for $\Omega_{\rm m}$. This reveals that clusters and voids (0-cycles and 2-cycles) are most informative for $\Omega_{\rm m}$, while $f_{\rm NL}^{\rm loc}$ is additionally informed by filaments (1-cycles).

}

\begin{document}
\maketitle
\flushbottom

\section{Introduction}

The large-scale structure (LSS) of the universe, traced by the distribution of galaxies, halos, filaments, and voids, encodes invaluable information about the underlying cosmological model, initial conditions, and the physical processes driving cosmic evolution. Cosmological parameter inference, which aims to extract this information, is central to advancing our understanding of the universe. Upcoming observational surveys such as SPHEREx \cite{SPHEREx:2014bgr}, Euclid \cite{Amendola:2016saw}, LSST \cite{Zhan:2017uwu} promise to significantly enhance our ability to probe the LSS at groundbreaking precision. However, fully leveraging the wealth of information contained in the LSS remains a major challenge, particularly in the nonlinear regime where gravitational collapse gives rise to intricate, strongly non-Gaussian features.

Traditional summary statistics, such as the two-point correlation function and its Fourier transform, the power spectrum, have long been the conventional tools for analyzing the LSS. These low-order statistics effectively capture the primary properties of the LSS, such as clustering and density fluctuations, and have been widely used to constrain parameters like the matter density $\Omega_{\rm m}$ and the amplitude of matter fluctuations $\sigma_8$ \cite{Ivanov:2019pdj,Zhang:2021yna,Ivanov:2019hqk,DES:2020ahh,DES:2020mlx,Vikhlinin:2008ym,rozo2009cosmological,SPT:2024qbr,Ghirardini:2024yni,eBOSS:2020yzd}. Higher-order statistics, such as the bispectrum, extend this framework by probing non-Gaussian features of the matter distribution \cite{Verde:2001sf,Gil-Marin:2014sta,DAmico:2019fhj,DAmico:2022osl,Ivanov:2023qzb,Colas:2019ret}. However, these methods often struggle to capture the full complexity of the nonlinear, small-scale regime. Moreover, the question of which summary statistics are most effective for analyzing non-Gaussian fields remains unresolved, as different approaches may capture complementary aspects of the underlying physics.

This effort is motivated by the goal of uncovering the fundamental physical processes shaping the universe. Of particular interest for primordial cosmology is the study of the deviation of initial metric fluctuations from a Gaussian distribution, i.e. primordial non-Gaussianity (PNG). However, gravitational evolution also sources non-Gaussianities, and local phenomena, such as galaxy bias, can produce signals that are difficult to distinguish from those originating in the early universe \cite{Biagetti:2019bnp}. This overlap creates significant challenges for traditional summary statistics, which are often unable to separate late-time effects from signatures of fundamental physics \cite{Cabass:2022ymb,Cabass:2022wjy,DAmico:2022gki}. Thus, novel methodologies are needed to disentangle these contributions and reliably identify the unique fingerprints of PNG, particularly on nonlinear scales where the interplay between these processes becomes most complex.

Recognizing this limitation, a range of alternative summary statistics have been proposed to extend the reach of LSS analyses into the nonlinear regime. Among them \cite{White:2016yhs,Jung:2024esv,Massara:2020pli,Marinucci:2024bdq,Massara:2022zrf,Cowell:2024wyl}, skew-spectra \cite{Hou:2024blc,Schmittfull:2020hoi}, Wavelet Scattering
 Transform \cite{Peron:2024xaw,Eickenberg:2022qvy,Valogiannis:2022xwu,Valogiannis:2023mxf}, one-point PDFs \cite{Uhlemann:2019gni,Friedrich:2019byw,Gould:2024sve}, Void Abundance \cite{DAmico:2010dwy,Kamionkowski:2008sr,Pisani:2019cvo}, k-nearest neighbors \cite{Banerjee:2020umh,Coulton:2023ouk}, Minkowski functionals \cite{Lippich:2020vpy,Liu:2023qrj,Jiang:2023nzz}. This is but a small sample of the many methods being developed in this very active area of research. Additionally, field-level inference, which proposes to perform inference directly on the entire density field bypassing the need for summary statistics, presents a promising approach \cite{SimBIG:2023ywd,Shao:2022mzk,deSanti:2023zzn,Anagnostidis:2022rbs,Chatterjee:2024eur}. By leveraging the full information contained in the data, field-level inference methods theoretically offer the greatest potential for maximizing the extraction of cosmological parameters. However, they require heavy machine learning machinery combined with extensive simulations. Such methods often yield results that are difficult to interpret and validate, particularly in understanding which field features are responsible for constraining the parameters of interest. In addition, they strongly rely on building reliable and ``accurate'' simulations, which might be a problematic bottleneck, especially when integrating astrophysical processes. These challenges are compounded by the significant computational demands of building such simulations. 
Another avenue is using forward modelling of the distribution of galaxies using the effective field theory of LSS \cite{Barreira:2021ukk,Nguyen:2024yth,Babic:2024wph,Kostic:2022vok}, but this is still limited to relatively large scales.  

This highlights the need to develop efficient and interpretable summary statistics that balance the high-information content of field-level analysis with the computational efficiency of traditional methods. In this context, a technique from topological data analysis (TDA), persistent homology, has emerged as a powerful tool for studying complex data structures. It is particularly well-suited for cosmology, where the LSS is organized into a hierarchical web of halos, filaments, and voids \cite{Wilding:2020oza}. By tracking the ``birth'' and ``death'' of topological features ---such as clusters, loops, and cavities--- across scales, persistent homology provides a unique, multi-scale description of the universe.

The outputs of persistent homology, known as persistence diagrams (PDs), can be transformed into persistence images (PIs), which summarize topological features in a format amenable to machine learning and statistical inference. While PIs could be directly incorporated into Bayesian inference frameworks, practical challenges such as estimating high-dimensional covariance matrices and capturing higher-order correlations, which would require an intractable number of simulations, have motivated the use of machine learning techniques. Neural-networks based inference, in particular, offer a robust approach for extracting patterns from PIs and addressing these challenges \cite{Yip:2023vud}. PIs are interpretable because the topological features they represent correspond to physically meaningful structures in the cosmic web.

Persistent homology has been successfully applied to a variety of problems in cosmology. Among the first implementations are~\cite{2010MNRAS.408.2163A,Sousbie2011,10.1093/mnras/stw2862}, with specific applications ranging from identifying primordial non-Gaussianity in N-body simulations \cite{Cole:2020gzt,Biagetti:2022qjl,Yip:2024hlz,Biagetti:2020skr,Feldbrugge_2019}, to analyzing weak lensing through cosmic shear simulations \cite{Heydenreich:2020hrr,Heydenreich:2022dci} and constraining the effects of massive neutrinos on the matter field \cite{Kanafi:2023twi}.


This paper is a continuation of \cite{Yip:2023vud}, which pioneered combining computational topology with machine learning in the context of cosmology. Using a convolutional neural network model, we map persistence images to cosmological parameters, enabling the extraction of information beyond that captured by traditional summary statistics. For benchmarking, we compare the performance of PIs with constraints derived from the power spectrum and bispectrum combined, offering a comprehensive evaluation of their relative information content.

We extend prior efforts in several key ways. First, we employ high-fidelity simulations from larger volumes and three independent datasets, ensuring robust and realistic parameter recovery. Second, we expand the parameter space to include both standard cosmological parameters and primordial non-Gaussianity amplitudes, providing a broader test of PIs' constraining power. In addition to predicting cosmological parameters, our models estimate associated uncertainties, offering a more complete characterization of the inference process. Finally, we enhance the interpretability of our results by employing feature-scoring methods, such as gradient-boosted trees, to identify the specific features of PIs that contribute most to parameter constraints. 


This paper is organized as follows. Section \ref{sec:ph} introduces the persistent homology framework and the construction of persistence images, emphasizing their relevance for LSS analysis. In Section \ref{sec:summarystats} we outline the dataset we use to compute our summary statistics built from the persistent homology pipeline. Additionally, we describe the building of the traditional power spectrum and bispectrum, computed from the halo catalogs used in this work. Section \ref{sec:modeltraining} describes the machine learning models used for parameter inference, including convolutional neural networks and gradient-boosted trees. Section \ref{sec:results} presents the main results, comparing the performance of PIs with traditional summary statistics in constraining cosmological parameters. Within this section, we also examine the sensitivity of PIs to individual cosmological parameters using feature attribution methods. Finally, Section \ref{sec:conclusions} discusses the broader implications of our findings and outlines future directions, such as integrating persistent homology with simulation-based inference frameworks and applying these methods to galaxy survey data.

\section{Topological data analysis and persistent homology}
\label{sec:ph}
Topological data analysis (TDA) is a collection of methods originating in algebraic topology, a branch of mathematics founded by Poincaré that studies the shape of data through its topological features. Central to TDA is the notion of homology, which captures topological features such as connected components, loops, and voids in a given space.

In particular, persistent homology is a tool that extracts and quantifies the persistence of these features, enabling the identification of robust, scale-independent topological signals in noisy or high-dimensional data. We refer the reader to~\cite{books/daglib/0025666,carlsson2021topological,wasserman2016topologicaldataanalysis} for comprehensive introductions to foundational developments in TDA.

Persistent homology can be applied to either discrete point sets or continuous fields, provided a notion of filtration is defined. Conceptually, a filtration is a family of nested sets parametrized by a filtration parameter (also called filtration time), denoted by $\nu$. This parameter typically reflects a notion of proximity in the data space. For each value of this geometric parameter, there is a set in the filtration in which topological features can be identified. Hence, as the parameter varies, topological features come into existence, evolve, and eventually die at various parameter values. Persistence statistics can then be built from this history of topological evolution.

\subsection{Persistent homology of the large-scale structure}
\begin{figure}
    \centering
        \includegraphics[width=0.5\textwidth, clip]{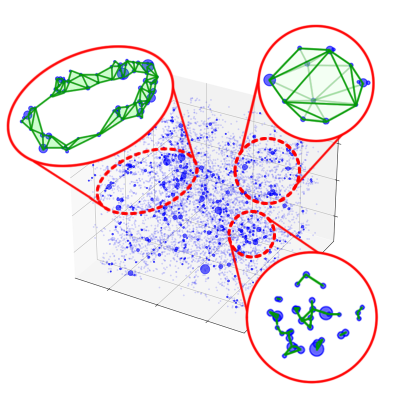}
        \caption{Cosmic structures are topological: Halo clusters (bottom right), filament loops (top left), and cosmic voids (top right) correspond to the 0-, 1-, and 2-cycles in topology, respectively.}
        \label{fig:cosmictopo}
\end{figure}

In the context of cosmology, this tool can be used to capture the multi-scale topological characteristics of the large-scale structure, roughly as a collection of clusters, loops, and voids distributed hierarchically in scale. These cosmological structures arise within the distribution of dark matter, which can be traced by dark matter halos, the hosts of visible galaxies. Hence, one expects that persistent homology applied to the spatial distribution of these halos probes these structures. To build an intuition, one can roughly interpret the large-scale structure, when expressed in terms of homology groups, as follows: high-density halo clusters extending into walls and filaments ($0$-cycles), loop-like filamentary bridges connecting matter concentrations ($1$-cycles), and fully enclosed cavities forming the central regions of cosmic voids ($2$-cycles). These topological features are not necessarily one-to-one mappings with visually defined cosmic structures, but rather  mathematically defined equivalence classes (cycles modulo boundaries) that reflect how matter is connected at multiple scales. Figure~\ref{fig:cosmictopo} offers a heuristic illustration of how these topological features may arise from halo distribution. The filtration values, which indicate the scales at which these features are born and die, encode geometric information that can be used to infer the underlying cosmological model.

\begin{figure}
    \centering
        \includegraphics[width=0.8\textwidth, clip]{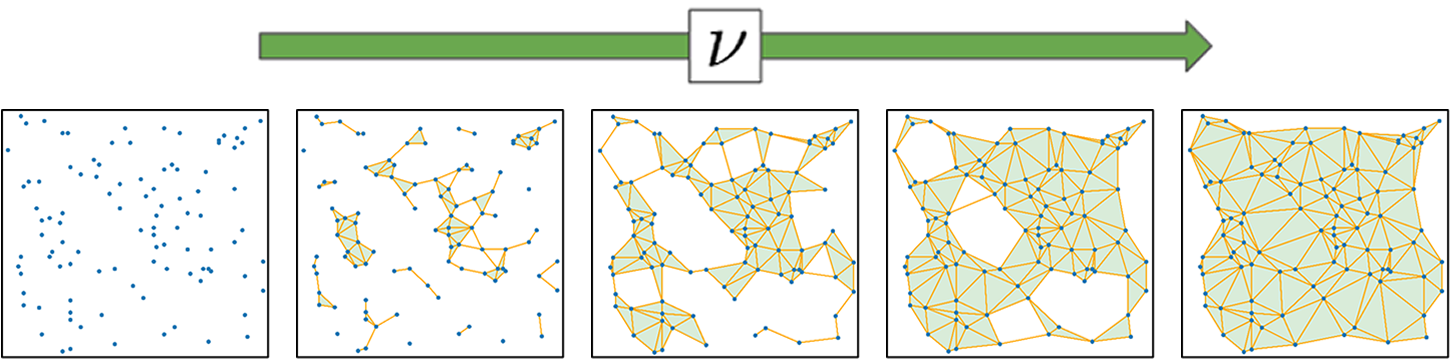}
        \caption{Example of a filtration for a point cloud in $2$D, as a collection of simplicial complexes parametrized by the filtration parameter or time $\nu$. As $\nu$ increases, $n$-simplices are subsequently added following a set of rules which can be flexibly designed. Topological features emerge and die throughout the filtration.}
        \label{fig:filtration}
\end{figure}

Let us walk through the implementation.\footnote{We refer the reader to \cite{Yip:2024hlz} for a more comprehensive and gentle introduction to the same implementation.} Given a point cloud of halo positions in $3$ dimensions, to define a filtration for it is to write down a set of rules that adds connections between the halos depending on the value of the non-negative\footnote{Hence the expression ``filtration time''.} filtration parameter $\nu$. Here, connections refer to the $n$-simplices where halos act as the vertices: the $1$-, $2$-, and $3$-simplices, which are edges, triangular faces, and solid tetrahedra respectively. An intuitive description of a filtration is that $\nu$ represents spatial distance, and a simplex is added if its size is equal to, or smaller than, $\nu$. Accordingly, more simplices are included as we dial $\nu$, and this collection of simplices, known as a simplicial complex, becomes increasingly sophisticated in its connectedness, eventually leading to successive emergences of topological features. In other words, the filtration is the family of these simplicial complexes at varying $\nu$'s (Figure \ref{fig:filtration})\footnote{The example shown is an $\alpha$-filtration, which is what the $\alpha$-DTM-$\ell$ filtration is based on.}. For this paper, the $n$-simplices are taken from the Delaunay triangulation of the point cloud, and we adopt the $\alpha$-DTM-$\ell$ filtration, which we explain in the next subsection. As a final note, formally known as a $p$-cycle, a topological feature is an equivalence class of boundaryless collections of $p$-simplices where each collection is not itself the boundary of any collection of $p+1$-simplices. Intuitively, these equivalent collections enclose the same ($p+1$)-dimensional ``hole'' that characterizes the topological feature.\footnote{In plain terms, a $0$-cycle is a collection of connected simplices, a $1$-cycle is a loop, and a $2$-cycle is an enclosed cavity.}

\subsection{Lightning review of $\alpha$-DTM-$\ell$ filtration}
The $\alpha$-DTM-$\ell$ filtration was first implemented on halos in \cite{Biagetti:2020skr}. It is based on the $\alpha$-filtration (or $\alpha$-shape)~\cite{10.1145/174462.156635} defined by $\alpha$-complexes~\cite{1056714}, which is well-established and widely implemented by libraries such as \texttt{GUDHI}~\cite{gudhi:urm}, which uses the \texttt{CGAL} library~\cite{cgal:eb-19a}. The $\alpha$-filtration has previously been applied to the large-scale structures, see~\cite{2010MNRAS.408.2163A,vandeweygaert2013alphabettimegaparsecuniverse,bermejo2024topologicalbiashaloestrace}.

While the underlying simplicial structure remains based on Delaunay triangulations, the key feature of the $\alpha$-DTM-$\ell$ filtration is that it uses the density-aware DTM (Distance-To-Measure) function to replace Euclidean distance in $\alpha$-filtration. There is a parameter, the number of nearest neighbors used $k$,\footnote{Not to be confused with the wavenumber $k$ associated with the power spectrum and bispectrum.} adjustable for extracting topological information at different scales \cite{Yip:2024hlz}. The set of rules defining the filtration is as follows:
\begin{enumerate}
    \item We begin (at $\nu=0$) with an empty simplicial complex, not even containing vertices (halos). A vertex $x$ is added at $\nu_x={\rm DTM}_x$, where
    \begin{equation}\label{eq:dtm}
        {\rm DTM}_x\equiv\sqrt{\frac{1}{k}\sum\limits_{X_i\in \mathcal{N}_k(x)}\|x-X_i\|^2}
    \end{equation}
    is the Distance-To-Measure function which quantifies the sparsity around the halo that $x$ corresponds to. Here $\mathcal{N}_k(x)$ is to $x$ the set of $k$-nearest neighbors in the given halo point cloud, and $\|a-b\|$ is the Euclidean distance between vertex $a$ and $b$. In other words, if the halo in question lies in a sparsely populated region, then ${\rm DTM}_x$ will be large, and $x$ is added at a late filtration time.
    
    \item An edge $\sigma_{x_1x_2}$ linking the vertices $x_1$ and $x_2$ is added if
    \begin{equation}\label{dx1x2}
        d_{x_1x_2}\equiv\|x_1-x_2\|\leq r_{x_1}(\nu)+r_{x_2}(\nu),
    \end{equation} where
    \begin{equation}\label{rx}
        r_x(\nu)\equiv\sqrt{\nu^2-{\rm DTM}_x^2}.
    \end{equation}
    Alternatively, we can determine the time $\nu_{\sigma_{x_1x_2}}$ at which the edge $\sigma_{x_1x_2}$ is added by solving eqs. (\ref{dx1x2}) and (\ref{rx}):
    \begin{equation}
        \nu_{\sigma_{x_1x_2}}=\sqrt{\frac{\left(\left({\rm DTM}_{x_1}+{\rm DTM}_{x_2}\right)^2+d_{x_1x_2}^2\right)\left(\left({\rm DTM}_{x_1}-{\rm DTM}_{x_2}\right)^2+d_{x_1x_2}^2\right)}{2d_{x_1x_2}}}.
    \end{equation}
    
    \item Higher-dimensional simplices ($2$- and $3$-simplices; triangles and tetrahedra) are added immediately when the necessary lower-dimensional faces (edges or triangules) are present. For example, if the edges $\sigma_{x_1x_2}$, $\sigma_{x_1x_3}$, and $\sigma_{x_2x_3}$ are added one after another, then the triangle $\sigma_{x_1x_2x_3}$ is also added at the same filtration time as $\sigma_{x_2x_3}$, i.e., $\nu_{\sigma_{x_1x_2x_3}}=\nu_{\sigma_{x_2x_3}}$.\footnote{Provided that the Delaunay triangulation of the given point cloud contains $\sigma_{x_1x_2x_3}$, which is not guaranteed for arbitrary combinations of $x_1$, $x_2$, and $x_3$.}
\end{enumerate}

One can visualize Rule 2 as placing around each vertex $x_i$ a growing sphere of radius $r_{x_i}(\nu)$, and the corresponding edge is added when two spheres touch or overlap. ${\rm DTM}_{x_i}$ impedes the growth of the sphere and delaying the inclusion of simplices involving $x_i$. As presented in \cite{Yip:2024hlz}, tuning $k$ effectively regulates the average volume within which the algorithm explores for the densest regions to populate with simplices. Therefore, when a larger $k$ is used, some regions are deemed no longer dense enough for early population, resulting in larger holes (i.e., longer-lived topological features) to emerge.\footnote{Watch filtrations with different $k$'s in action: \url{https://youtu.be/_phgkiZmY0c}.} Intuitively, smaller values of $k$ result in finer-scale, more locally sensitive filtrations, potentially capturing smaller features. In contrast, larger $k$ values smooth out local density fluctuations, emphasizing more global topological features. In other words, we can vary $k$ to change the scale at which topological information is extracted.

\subsection{Persistence outputs: diagrams and images}
\begin{figure}
    \centering
        \includegraphics[width=1\textwidth, clip]{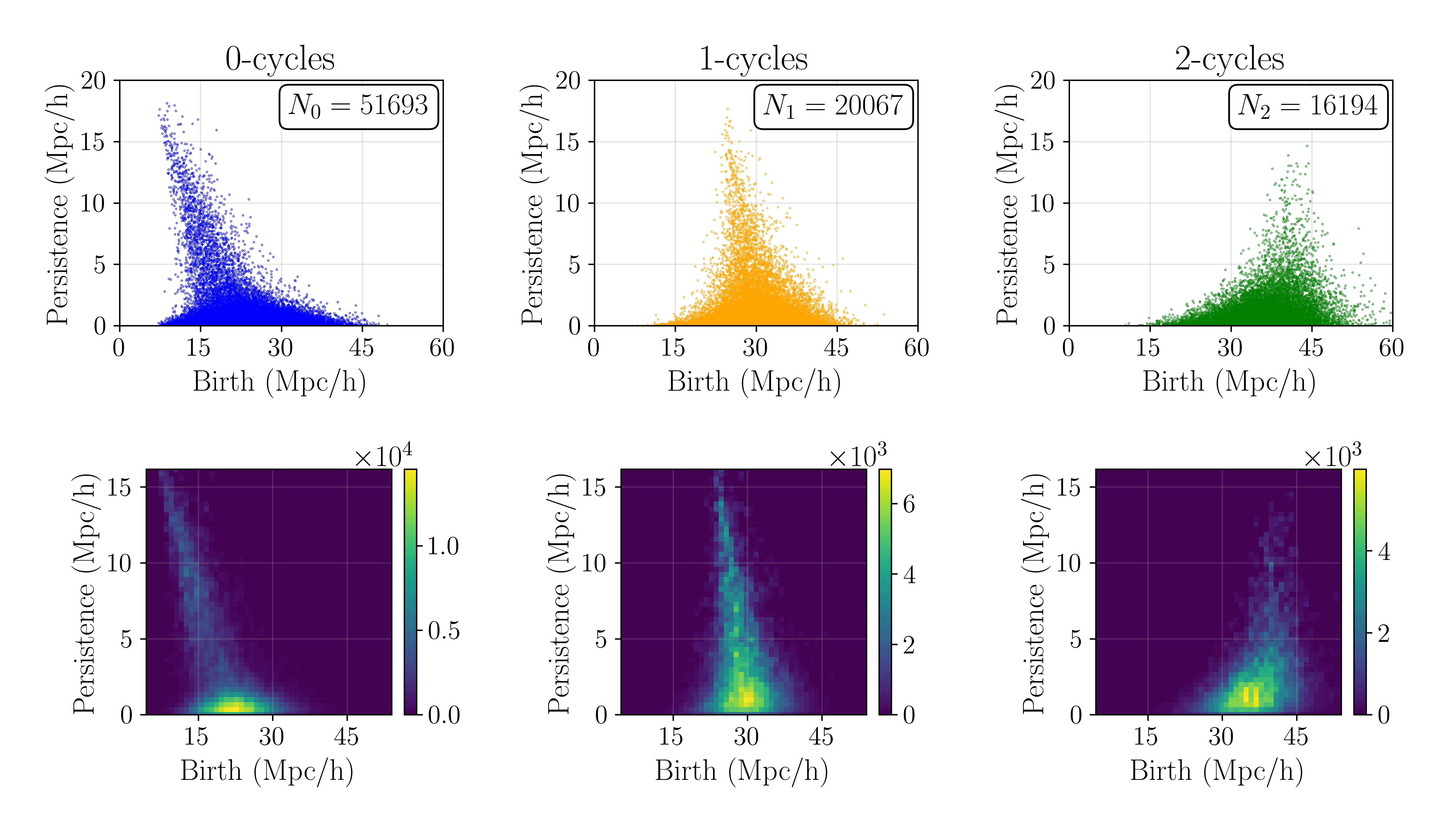}
        \caption{\emph{Top Panel:} Persistence diagrams of $0$-, $1$-, and $2$-cycles from the filtration of a \quijote halo catalog with $k=15$ at the fiducial cosmology. $N_p$ is the total number of $p$-cycles in each diagram, i.e., the total number of $p$-cycles that once existed in the filtration. \emph{Bottom Panel:} Corresponding persistence images from ``pixelating'' the diagrams.}
        \label{fig:pdpi}
\end{figure}

With the filtration defined, we now have a simplicial complex that evolves with the filtration parameter, or the filtration time, $\nu$. Topological features (holes) come into existence, persist, and are eventually trivialized (i.e., filled in by simplices). Using $\nu$ as the handle, we can quantify this topological history by tracking the life of every feature that has ever existed. Precisely, each feature is characterized by a tuple $(\nu_{\rm birth}, \nu_{\rm persist})$, where $\nu_{\rm birth}$ is the filtration time at which the feature is formed, and $\nu_{\rm persist}=\nu_{\rm death}-\nu_{\rm birth}$ is the duration of its existence prior to trivialization. To sum up, the primary outputs of a persistent homology computation are lists of $(\nu_{\rm birth}, \nu_{\rm persist})$ pairs, and in our $3$-dimensional application there are $3$ such lists for the $0$-, $1$-, and $2$-cycles. Each of these lists is often presented as \emph{persistence diagrams} and \emph{persistence images} (Figure \ref{fig:pdpi}):

\paragraph{Persistence diagrams} We plot all the cycles that ever existed in the filtration in the $\nu_{\rm persist}$-$\nu_{\rm birth}$ plane, generating persistence diagrams. There are $3$ persistence diagrams in our application, one for each of the $3$ homological dimensions.

\paragraph{Persistence images} Fully specifying an arbitrary persistence diagram requires $2\times($number of cycles$)$ values. In most scenarios, one would wish to work with a data vector of a fixed size. To this end, the conventional way of vectorizing a persistence diagram is to ``pixelate'' it into a persistence image. Specifically, we bin in two dimensions the $\nu_{\rm persist}$-$\nu_{\rm birth}$ plane, by assigning a smoothing kernel to each point in the persistence diagram. Then we sum up all kernel contributions in each bin.

Persistence images have proven useful in large-scale structures analysis, such as~\cite{10.1093/mnras/stw2862}. There are other vectorizations that condense the information in a persistence diagram even further. An example is the concatenation of histograms of $\nu_{\rm birth}$ and $\nu_{\rm death}$ values, which is used in \cite{Biagetti:2022qjl,Yip:2024hlz}, where a Gaussian likelihood is assumed in Fisher forecast setups. In this work, we use neural network models for inference. As a result, we are less constrained and need to summarize the persistence diagrams only minimally, allowing for a controlled amount of information loss.

\section{Summary statistics from simulations}
\label{sec:summarystats}

The entirety of our analysis is simulation-based. In this section, we provide a detailed description of how the persistence images, as well as the joint power spectrum and bispectrum statistic, are measured from the simulations.

\subsection{Halo catalogs}
\begin{figure}
    \centering
        \includegraphics[width=1\textwidth, clip]{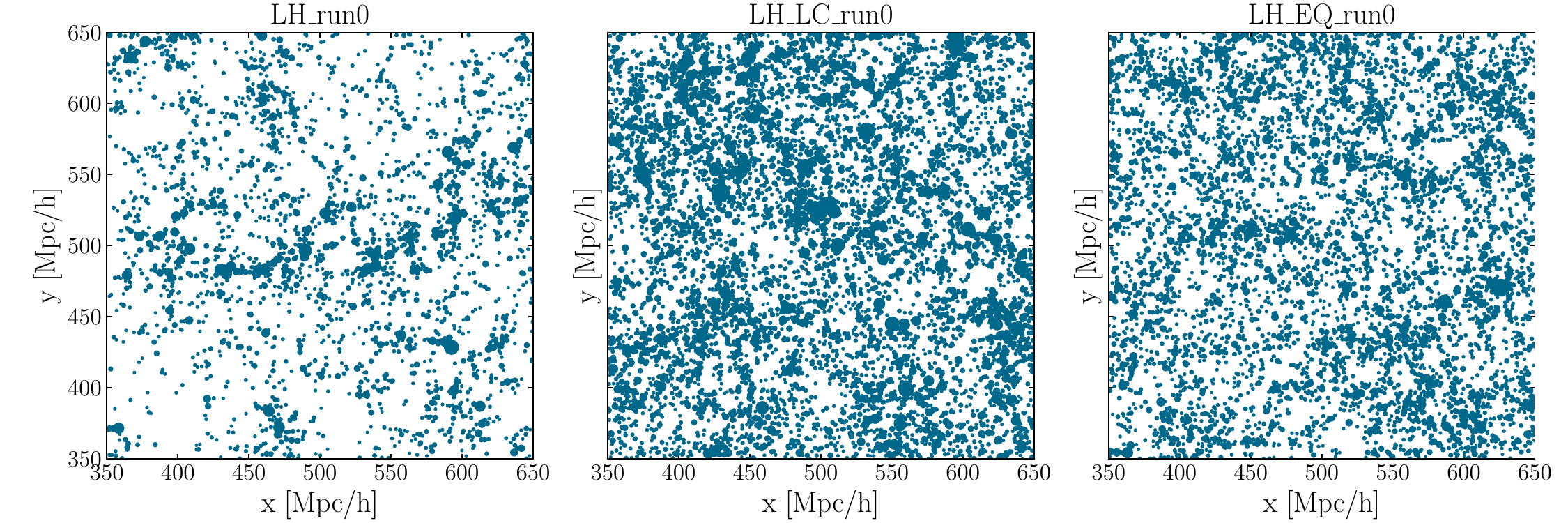}
        \caption{This figure shows a slice of thickness $250\,\mathrm{Mpc}/h$ along the $z$-direction, centered at $500\,\mathrm{Mpc}/h$. Point sizes are proportional to halo mass. All panels share the same initial random seed but differ in cosmological parameters. The \textit{left panel} corresponds to a realization from the LH dataset with $\Omega_{\rm m} = 0.1755$, $\Omega_{\rm b} = 0.06681$, $h = 0.7737$, $n_{\rm s} = 0.8849$, and $\sigma_8 = 0.6641$; the \textit{middle panel} shows the same seed from the LH\_$f_\mathrm{NL}^\mathrm{local}$ dataset, with $f_\mathrm{NL}^{\mathrm{local}} = -98.7$ and fiducial cosmological parameters. The \textit{right panel} shows the LH\_$f_\mathrm{NL}^\mathrm{equi}$ dataset, also with the same seed, adopting $f_\mathrm{NL}^{\mathrm{equi}} = 201$, $\Omega_{\rm m} = 0.3330$, $h = 0.5966$, $n_{\rm s} = 1.0726$, and $\sigma_8 = 0.7110$.}
        \label{fig:halos_slices}
\end{figure}

We use the halo catalogs obtained from the N-body simulations in the \quijote and \quijotepng suites \cite{Villaescusa-Navarro:2019bje,Coulton:2022qbc} (hereafter collectively referred to as \quijote). In each of these simulations, $512^3$ dark matter particles are evolved using \texttt{GADGET-III} within a cosmological volume of $1\,($Gpc$/h)^3$, with initial conditions at redshift $z=127$ generated by the public \texttt{2LPTIC} code. Dark matter halos are identified using the Friends-of-Friends (FoF) algorithm, and we utilize halo catalogs at $z=0.5$, which are comparable to galaxies observed in surveys such as the BOSS CMASS sample. We also exclude halos that are composed of fewer than $50$ particles, meaning that halos involved in our analysis have a minimum mass of $M_\text{min}=3.28\times10^{13}M_{\odot}$. We show in Fig.~\ref{fig:halos_slices} a visual comparison of halo spatial structures and clustering morphologies driven by different cosmological parameters, using fixed initial conditions. For further details on the simulation suite, we refer the reader to existing studies of the \quijotepng simulations~\cite{Jung:2023kjh,Jung:2024esv}, which explicitly analyze the halo mass function with power spectrum and bispectrum statistics.

The training, validation, and testing datasets of summary statistics for our neural network inference models are measured from the Latin-hypercube (LH) subset of the \quijote simulations.\footnote{\url{https://quijote-simulations.readthedocs.io/en/latest/LH.html}} Each simulation in an LH has a cosmology defined by cosmological parameters, including primordial non-Gaussianity amplitudes, which are sampled within specified ranges. We utilize $3$ of the available LHs: LH, LH\_$\fnlloc$, and LH\_$\fnleq$. For the fiducial parameter recovery tests, we employ the $15{,}000$ fiducial realizations from the main suite. See table \ref{tab:LHs} for a summary.

Since observations are carried out in redshift space, prior to measurements, we convert the real-space positions of the halos to redshift-space positions. In the distant-observer approximation, the conversion is
\begin{equation}
\boldsymbol{x}\rightarrow\boldsymbol{x}+\frac{\boldsymbol{v}\cdot\hat{\boldsymbol{n}}}{a(z)H(z)}\hat{\boldsymbol{n}},
\end{equation}
where $\boldsymbol{x}$ is a halo's real-space position, $\boldsymbol{v}$ is the halo's peculiar velocity, $\hat{\boldsymbol{n}}$ is the line-of-sight, $a(z=0.5)=(1+0.5)^{-1}$ is the scale factor, and the Hubble parameter $H(z)=100\sqrt{\Omega_{\rm m}a^{-3}+\Omega_{\Lambda}a^{-3(1+w)}}=100\sqrt{\Omega_{\rm m}a^{-3}+(1-\Omega_{\rm m})}$ in $({\rm km}/{\rm s})(h/{\rm Mpc})$ depends on $\Omega_{\rm m}$ of the catalog's cosmology. We process each catalog along the $z$-axis, i.e., taking $\hat{\boldsymbol{n}}=\hat{\boldsymbol{z}}$.


\begin{table}
    \centering
    \scriptsize
    \begin{tabular}{|l"c"c|c|c|c|c|c|c|}
        \hline
        \textbf{Category} & \makecell{\tiny\textbf{Number of}\\\tiny\textbf{realizations}} & $\Omega_{\rm m}$ & $\Omega_{\rm b}$ & $h$ & $n_{\rm s}$ & $\sigma_8$ & $f_{\rm NL}^{\rm loc}$ & $f_{\rm NL}^{\rm equi}$ \\
        \thickhline
        Fiducial & $15000$ & $0.3175$ & $0.049$ & $0.6711$ & $0.9624$ & $0.834$ & $0$ & $0$ \\    
         \hline
        LH & $2000$ & $[0.1,0.5]$ & $[0.03,0.07]$ & $[0.5,0.9]$ & $[0.8,1.2]$ & $[0.6,1.0]$ & $0$ & $0$ \\
         \hline
        LH\_$\fnleq$ & $1000$ & $[0.1,0.5]$ & $0.049$ & $[0.5,0.9]$ & $[0.8,1.2]$ & $[0.6,1.0]$ & $0$ & $[-600,600]$ \\
         \hline
        LH\_$\fnlloc$ & $1000$ & $0.3175$ & $0.049$ & $0.6711$ & $0.9624$ & $0.834$ & $[-300,300]$ & $0$ \\
         \hline
    \end{tabular}
    \caption{Simulation sets from \quijote and \quijotepng used in this work. The different LH (Latin-hypercube) realizations are used for the training and evaluation of the models, while the Fiducial simulations are used exclusively for the fiducial parameter recovery tests.\label{tab:LHs}}
\end{table}

\subsection{Persistence images}

As described in section \ref{sec:ph}, we vary the nearest-neighbor parameter $k$ in the $\alpha$-DTM-$\ell$ filtration of our persistent homology pipeline. We choose $k\in\{1,5,15,30,60,100\}$, which was shown in \cite{Yip:2023vud} to be an informative selection for \quijote's simulation volume and halo density; the persistence statistic derived from each value is expected to contain a reasonable amount of independent information. In summary, for a single halo catalog we have $(3$ homological dimensions$)\times(6$ $k$ values$)=18$ images.

We set the resolution of each persistence image to be $128\times128$ pixels, which provides a good trade-off between granularity and dimensionality given the size of our datasets. The bounds, i.e., minimum and maximum birth values, and maximum persistence values covered by an image are shared across images for the same homological dimension and value of $k$. We find the said bounds by surveying the LH set of simulations: for each realization we take the $1$st and $99$th percentile of all birth values and the $99$th percentile of all persistence values, then we pool these values from all realizations and use the $1$st/$99$th percentiles as the bounds. The first step removes outliers in each realization, while the second removes outliers within the simulation set. With this setup, we set for kernel density estimation the bandwidth parameter in the \texttt{KernelDensity} function in the \texttt{scikit-learn} library to be $5\times($persistence per pixel$)$, where persistence per pixel is defined as the persistence bound divided by $128$ (the 1D resolution).

\subsection{Power spectrum and bispectrum}
Following \cite{Yip:2024hlz}, we compute the redshift-space monopole and quadrupole components of the halo-halo power spectrum and the monopole of the bispectrum using the publicly available code PBI4.\footnote{Available at: \url{https://github.com/matteobiagetti/pbi4}} The bin widths are defined as $\Delta k = 2 k_f$, with the first bin centered at $2k_f$, where $k_f = 0.006 \, h \text{Mpc}^{-1}$ is the fundamental frequency of the simulation box. This binning procedure extends up to $k_{\text{max}} = 0.3 \, h \text{Mpc}^{-1}$ beyond which shot noise dominates the signal, resulting in a total of 24 bins for each power spectrum component and 1522 triangular bins for the bispectrum monopole alone. To reduce aliasing effects, we implement the interlacing scheme of \cite{Sefusatti:2015aex}, combined with a fourth-order interpolation scheme, to compute the density contrast on a Fourier grid with 144 bins.

For both the power spectrum and the bispectrum monopole, we subtract a pure Poisson shot noise term, given by $1/\bar{n}$ for the power spectrum, and $1/\bar{n} \left(P(k_1) + P(k_2) + P(k_3)\right) + 1/\bar{n}^2$ for the bispectrum, where $\bar{n}$ is the halo density of the catalog.

\section{Model architectures, training, and hyperparameter optimization}
\label{sec:modeltraining}

In this section, we present the different models used in our analysis, as well as their training and optimization strategies for predicting cosmological parameters from persistence images (PIs), power spectrum and bispectrum (PS/BS), and both combined.
\\


\subsection{Convolutional Neural Networks for persistence images}

Our generic \cnn architecture combines two types of convolutional blocks. The first type consists of a Conv2D layer with a kernel size of 3, activated by LeakyReLU, followed by a max-pooling operation with a kernel size of 2. The second type is similar but removes the max-pooling operation. The first block is applied several times, followed by a couple of applications of the second block. The data is flattened after the convolutional layers to produce a one-dimensional vector and passed through a dropout layer before being fed into a fully connected output layer. This final layer predicts both the mean and standard deviation, as described below in section \ref{sec:loss}. The specific number of blocks and channels, as well as the dropout rates, weight decay, and learning rate were optimized through a hyperparameter search for individual datasets (i.e. LHs), as described in section \ref{subsec:train-hpsearch}. 

We apply a 2D batch normalization layer directly to the input images for the specific case of the LH\_$\fnlloc$ dataset. In contrast, for the other Latin Hypercubes, we include the 2D batch normalization after every convolutional layer within each block, as we observed that this gives the best results.

\subsection{Multi-Layer Perceptron for joint power spectrum and bispectrum}

For the joint power spectrum and bispectrum statistics, we use a Multi-Layer Perceptron (\mlp) model. The input consists of the concatenation of the total power spectrum and bispectrum statistics, denoted as $\left[P_0, P_2, B_0\right]$.

The \mlp architecture applies a batch normalization for the input to ensure consistent scales across features. This is followed by several hidden layers, each composed of a linear layer with a LeakyReLU activation function and a dropout for regularization. The output layer is a linear layer that outputs the mean and the standard deviation for each parameter. We performed hyperparameter optimization on the learning rate, weight decay, number of neurons, layers and dropout rates.

While this general architecture was applied consistently to every Latin hypercube dataset, we explored  various architectural modifications to assess their potential for improvement. For instance, we tested adding batch normalization after every hidden layer, as well as separating the power spectrum and bispectrum statistics into two distinct branches. In this latter approach, the power spectrum components $\left[P_0, P_2\right]$ and the bispectrum $\left[B_0\right]$ were processed independently using separate {\mlp}s. The outputs of these branches were then concatenated into a single vector and passed through a final output layer for regression. However, none of these variations yielded significant improvements over the simpler design described above.


\subsection{Joint PI-PS-BS architecture}

To integrate persistence images (PIs) with power spectrum and bispectrum (PS/BS) statistics, we use a hybrid architecture that combines a convolutional neural network (\cnn) for the PI data and a multi-layer perceptron (\mlp) for the PS/BS data, referred to as \hybrid in the remainder of the paper.

The \mlp branch processes the PS/BS input data using the previously described \mlp architecture, excluding its output layer to focus on feature extraction. Similarly, the PI branch processes its input data through a \cnn model, also excluding its output layer. The representations extracted from both branches are then flattened and concatenated into a single feature vector. This combined representation is passed through a final regression layer, composed of fully connected layers with LeakyReLU activation functions and dropout, which predict the means and standard deviations for each cosmological parameter.

To improve training efficiency and performance, \hybrid is initialized with pre-trained weights from the independently trained \mlp and \cnn models. For the output fully connected layers, we performed hyperparameter tuning on the number of layers, number of neurons, weight decay, and learning rate.

\subsection{Training and Hyperparameter Optimization of Neural Network Models}
\label{subsec:train-hpsearch}

To optimize model performance across datasets, we performed hyperparameter search using Bayesian optimization through the \texttt{gp\_minimize} function from the \texttt{scikit-optimize} library.\footnote{\url{scikit-optimize.github.io/stable/modules/generated/skopt.gp_minimize.html}} The search explored configurations such as the number of layers, dropout rates, and learning rates. Initial trials sampled hyperparameters randomly, followed by refined sampling through Bayesian optimization.

For training, we used the Adam optimizer without a learning rate scheduler. Weights were initialized using the Kaiming normal distribution for LeakyReLU activations, and biases were initialized to zero. We used early stopping with a patience of 100 epochs to prevent overfitting. For all models, we used a fixed batch size of 32 and minimized the validation loss. The search process started with 10 random trials to explore the hyperparameter space, followed by 40 additional trials using Bayesian optimization. 

The hyperparameter space included the following configurations: for the \mlp, the number of layers was defined within the range $[2, 6]$, and the hidden layer sizes were chosen between the values $[32, 64, 128, 256, 512]$. For \cnn, the number of max-pooling Conv2D blocks was set in the range of $[3, 7]$, and the number of Conv2D blocks was defined within the range of $[1, 4]$. Furthermore, the number of channels for the Conv2D blocks was selected from $[8, 16, 32, 64, 128]$. For the \hybrid, the number of layers in the final output layer ranged from $[1, 8]$, with layer sizes selected from $[8, 16, 32, 64, 128, 256, 512]$. Dropout rates were chosen from $[0.3, 0.4, 0.5]$. The learning rate was sampled from a log-uniform distribution in the range $[10^{-4}, 10^{-2}]$, and weight decay was sampled log-uniformly from $[10^{-5}, 10^{-1}]$ for all models.

The data split was consistent across all models to ensure uniform conditions. For the LH dataset, the split consisted of 1600 training samples, 200 validation samples, and 200 test samples. For the LH\_$\fnlloc$ and LH\_$\fnleq$ datasets, we used 600 training samples, with 200 each for validation and testing. The same random split was applied across all models for training, validation, and test sets. This ensures that each model was evaluated on identical cosmologies in the test set, allowing direct performance comparison across architectures (and thus summary statistics).

\subsection{Likelihood-free parameter inference}
\label{sec:loss}

For parameter inference, we do not compute the full posteriors of the parameters $p(\theta_i|\bm{X})$,
where $\theta_i$'s represent the parameters we aim to recover. These are inferred from the summary statistic $\bm{X}$, which is derived from the halo catalog in the Latin hypercube dataset. Instead, our architecture predicts two values for each parameter: the mean $\hat{\mu}_i$ and the variance $\hat{\sigma}_i$ of the marginal posterior distribution. We use a custom loss function based on the logarithmic moment network methodology \cite{Jeffrey:2020itg}, which has recently gained attention for estimating model errors when a likelihood function is unavailable, as implemented in \cite{Villaescusa-Navarro:2021pkb,Villaescusa-Navarro:2021cni,Villanueva-Domingo:2021dun,Wang:2022zpv,Villanueva-Domingo:2022rvn,Perez:2022nlv,Villaescusa-Navarro:2022twv,deSanti:2023zzn,Chawak:2023bil,deSanti:2023rsw,Jung:2024esv,Gondhalekar:2024iqm}. The specific form of the loss function is as follows:
\begin{equation}\label{eq:loss}
    \mathcal{L} = \sum_{i \in \text{params}} \log \left( \sum_{j \in \text{batch}} (\theta_{i,j} - \hat{\mu}_{i,j})^2 \right) + \sum_{i \in \text{params}} \log \left( \sum_{j \in \text{batch}} \left( (\theta_{i,j} - \hat{\mu}_{i,j})^2 - \hat{\sigma}_{i,j}^2 \right)^2 \right)\,,
\end{equation}
where $\theta_{i,j}$ is the true value of the $i$-th parameter for the $j$-th sample in the batch, and $\hat{\mu}_{i,j}$ and $\hat{\sigma}_{i,j}$ denote the predicted mean and variance, respectively. The logarithmic terms ensure that both the mean and variance are similarly weighted across all parameters, preventing the loss function from being dominated by the least accurately estimated parameters.

We also tested a simple Mean Squared Error (MSE) loss function, given by $\mathcal{L} =  \frac{1}{N}\sum_{i \in \text{params}} (\theta_{i} - \hat{\mu}_{i})^2 $ to estimate the posterior means. In terms of architecture and results, the performance was comparable.

\subsection{Gradient Boosted Trees}

Although neural network-based methods are powerful and flexible methods to learn from complex data such as (natural) images or text, they can suffer from shortcomings on tabular data, where tree-based methods often outperform them in medium-sized datasets \cite{grinsztajn2022tree, Lazanu:2021tdl}. Additionally, tree-based methods seem more robust to uninformative features. While we refer to our Persistent Homology-based summary statistic as a Persistence \textit{Image}, it can actually be considered as tabular data as well: the location of a specific pixel has a meaning (birth scale and persistence). Power spectrum and bispectrum are also tabular data. Since we are in a relatively small dataset regime, where neural networks could overfit or require extreme regularization, we propose to use Gradient Boosted Trees (GBT) as an additional method to evaluate the predictive power of the summary statistics considered here and their robustness (or not) across types of methods (in our specific regime). We favor GBT over Random Forest as they have been shown to perform better for a comparable computational cost.

Another interesting advantage of decision tree-based models like \xgboost is their ability to evaluate feature importance, which can provide insights into how the model processes the features to predict. 
In Section \ref{sec:featuresimportance}, we provide visualizations and analysis of the feature importance on some of our datasets.



We train GBT using the \xgboost library \cite{Chen_2016} for each data combination. When persistence images are included, the pixel values are flattened into a single array. Datasets are split as for the neural nets. We perform grid-search with cross-validation to find the hyperparameters (learning rate, maximum depth, number of estimators, subsampling, max child weight) that best fit the training data. We find that models with very shallow trees are generally preferred (with a maximum depth of five). The reason may be that this avoids overfitting.
We minimize the RMSE to train these models and perform model selection on the validation. Since there is no off-the-shelf way to perform simulation-based inference techniques with boosted trees, we do not estimate the variance of each cosmological parameter. Therefore, these models are evaluated using only their RMSE and $R^2$ scores on the test set, and not the $\chi^2$.

It is valuable to note that the training of each model can be performed in a few minutes even in modest CPUs, which is significantly faster than neural-based approaches. However, they also require significantly more RAM.

\section{Results}
\label{sec:results}

We now evaluate the performance of the different methods and summary statistics on the different datasets (LHs) considered. We note that each model was trained and validated on a separate subset of the dataset, and no significant difference between training and validation losses was observed, indicating that the networks generalized well without overfitting.

\subsection{Evaluation Metrics}

To evaluate the performance of each model, we use several metrics to quantify the accuracy of the predicted means and variances for each cosmological parameter. These metrics include the coefficient of determination ($R^2$ score), the chi-squared statistic ($\chi^2$), and the root-mean-square deviation (RMSE).

\paragraph{$R^2$ score:} Measures the goodness of fit for the predicted mean, and therefore how well $\hat{\mu}_i$ tends to the true value $\theta_i$ for each parameter. It is given by
\begin{equation}
    R^2(\theta,\hat{\mu}) = 1 - \frac{\sum_i (\theta_i - \hat{\mu}_i)^2}{\sum_i (\theta_i - \bar{\theta})^2}\,,
\end{equation}
where the $\theta_i$ is the true parameter for the $i$-th sample, $\hat{\mu}_i$ is the predicted mean from the network model, and $\bar{\theta}$ is the average of the true parameter over the entire test set. The best score is reached at $R^2=1$, indicating a nearly perfect prediction, while $R^2 = 0$ means the prediction is no better than using the average value. Notice that $R^2$ can be negative meaning that the model performs worse than just predicting the mean.

\paragraph{$\chi^2$ score:} Evaluates how well the predicted variance $\hat{\sigma}$ matches the spread of the data. It is defined as
\begin{equation}
    \chi^2(\theta, \hat{\mu}, \hat{\sigma}) = \frac{1}{N} \sum_i \frac{(\theta_i - \hat{\mu}_i)^2}{\hat{\sigma}^2_i},
\end{equation}
where $N$ is the number of samples in the test set, $\theta_i$ is the true parameter value, $(\hat{\mu}_i,\hat{\sigma}_i)$ is the predicted mean and variance output by the model. A $\chi^2$ score close to 1 indicates that the predicted uncertainties are well-calibrated. If $\chi^2$ is significantly greater than $1$, the model underestimates the uncertainties. On the other hand, if it is significantly less than 1, the model overestimates them.

\paragraph{RMSE:} Quantifies the discrepancy between the predicted values generated by a model and the actual ground truth values, providing a comprehensive measure of the model's predictive performance relative to the true parameter values. A lower RMSE indicates a superior fit of the model to the data, thereby suggesting better predictive accuracy. It is defined as
\begin{equation}
    {\rm RMSE} = \sqrt{\frac{1}{N} \sum_i (\theta_i - \hat{\mu}_i)^2}.
\end{equation}

\subsection{Varying cosmologies}

\begin{figure}[]
    \centering
    \caption*{Persistence Image (PI) - LH dataset}
    \includegraphics[width=1\textwidth]{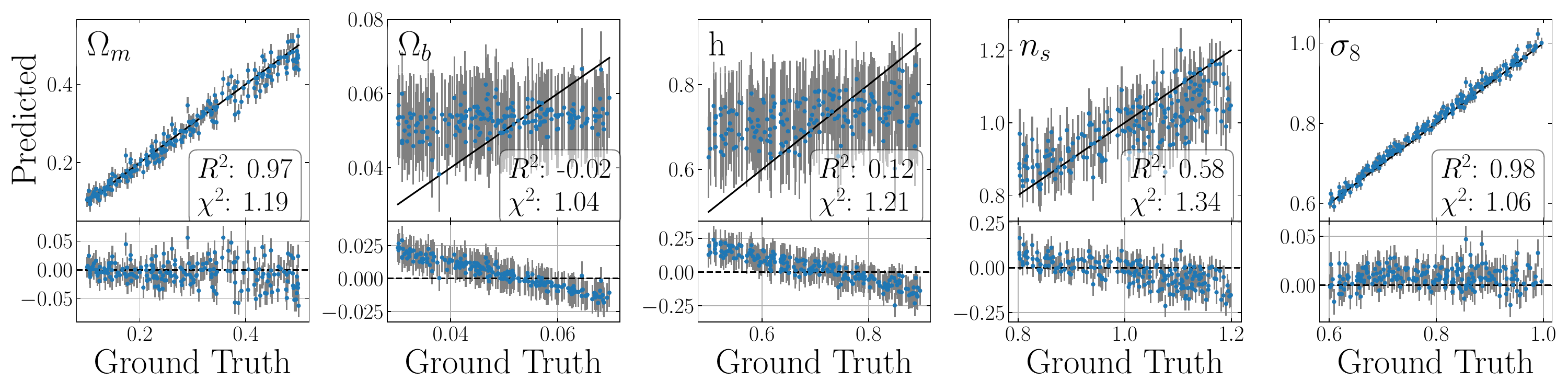}
    \caption*{Power Spectrum and Bispectrum (PSBS) - LH dataset}
    \includegraphics[width=1\textwidth]{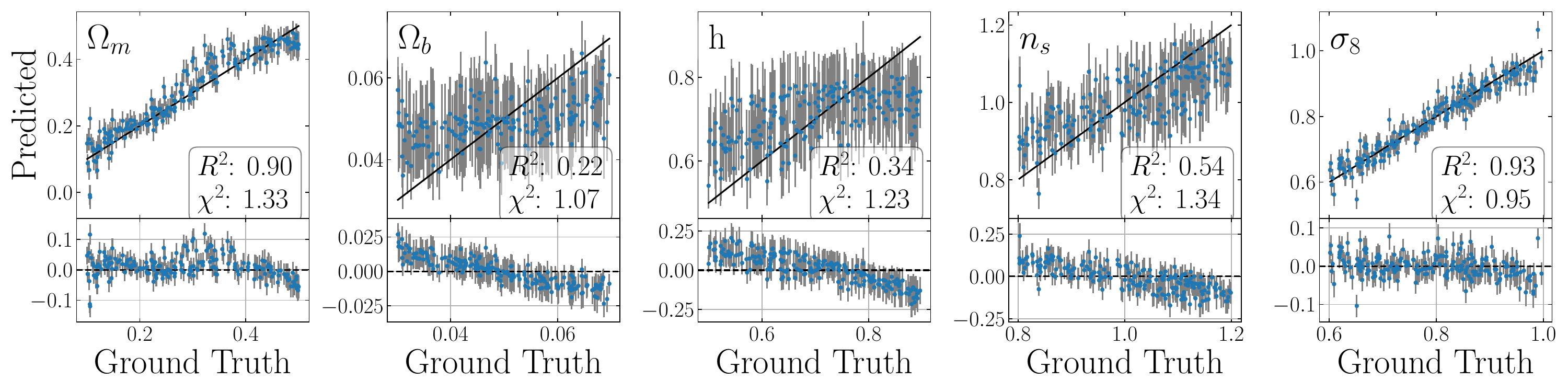}
    \caption*{Joint PI+PSBS - LH dataset}
    \includegraphics[width=1\textwidth]{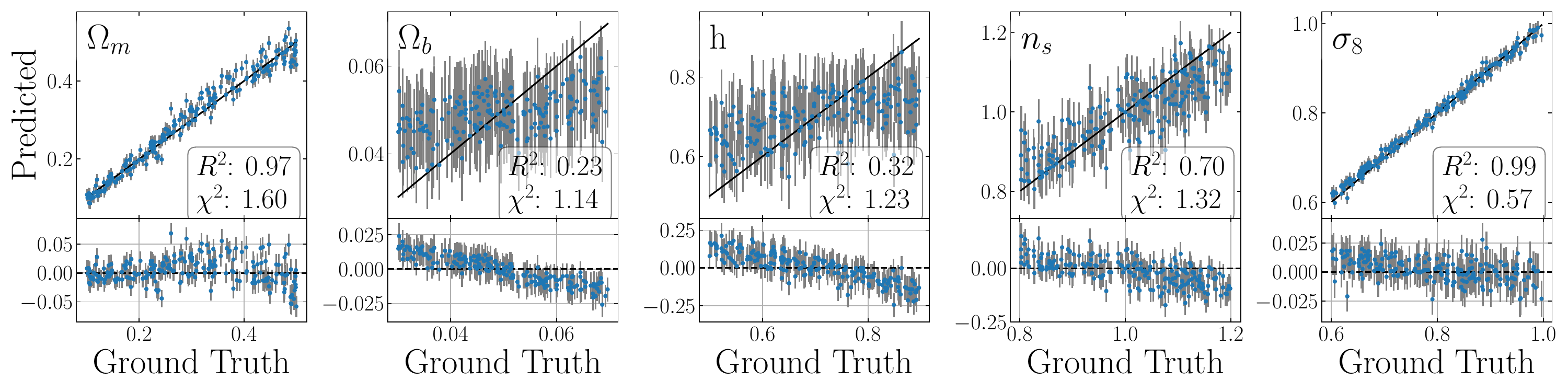}
    
    \caption{Model performance on predictions for all 200 test points using the best-performing models trained on the PI, PSBS, and PI-PSBS datasets. The black diagonal line represents a perfect match between predicted and true values. The error bars indicate the predicted standard deviation output by the network. The residuals are shown in the smaller lower panels. The dataset is the \quijote Latin Hypercube (LH) at redshift $z=0.5$.}
    \label{fig:main_plot}
\end{figure}

We provide in Fig.~\ref{fig:main_plot} the predicted mean posteriors along with their associated standard deviation (std) for all 200 test points in the LH dataset, for the three summary statistics (PI, PSBS, PI-PSBS) and their respective neural architectures. The predicted mean values from the network are plotted on the y-axis and the true (target) values are plotted on the x-axis. The black diagonal line represents a perfect match between predicted and true values. The error bars illustrate standard deviation estimated by the model for each prediction. Additionally, the smaller lower panels show the residuals between the predicted posterior means and the ground truth values, with the black line indicating a perfect match between the two. We see that for parameters which are well estimated, that is $\Omega_{\rm m}$ and $\sigma_8$, the predicted values and variances give an accurate statistical description of the model performance. On the other hand, for paramters that are poorly estimated, that is $\Omega_{\rm b}$ and $h$ the models are picking the mean among realizations, and though this is somewhat accounted for in the large error bars, there is a large bias visible in the residuals. For $n_{\rm s}$ the models capture some of the trend, assigning large error bars and a large bias. We also observe that none of these models have large outliers, but there is some slight but noticeable bias in the prediction of $\Omega_{\rm m}$ when using the PSBS data.

To quantify the performance of these models, we summarize the metrics---Root Mean Squared Error (RMSE), $R^2$, and $\chi^2$ in the following tables:  Table~\ref{tab:main_metrics} for the main Latin Hypercube (LH) dataset, Table~\ref{tab:fnl_loc_metrics} for the LH\_$\fnlloc$ dataset, and Table~\ref{tab:fnl_eq_metrics} for the LH\_$\fnleq$ dataset.


\hspace*{-1cm}
\begin{table}[h!]
\hspace{-1cm}
    \small{
    \begin{tabular}{|c || l |c c c c c|}
    \hline
    \textbf{Data}    & Metric  & $\Omega_{\rm m}$   & $\Omega_{\rm b}$  & $h$   & $n_{\rm s}$   & $\sigma_8$  \\[0.5ex]  \hline\hline 
    \multirow{3}{*}{\makecell[t]{PI \\ 
    (\textbf{\cnn})}}
                      & RMSE            & 0.025 $\pm$ 0.002 & 0.011 $\pm$ 0.001 & 0.11 $\pm$ 0.001 & 0.074 $\pm$ 0.006 & 0.012 $\pm$ 0.001      \\ 
    
                      & $\chi^2$        & 1.31 $\pm$ 0.20 & 1.04 $\pm$ 0.09 & 1.12 $\pm$ 0.06 & 1.14 $\pm$ 0.19 & 0.77 $\pm$ 0.12      \\ 
                      & $R^2$           & 0.96 $\pm$ 0.01 & -0.064 $\pm$ 0.09 & 0.14 $\pm$ 0.02 & 0.59 $\pm$ 0.06 & 0.99 $\pm$ 0.01  \\ \hline

    \multirow{2}{*}{\makecell[t]{PI \\ 
    (\textbf{\gbt})}}
                      & RMSE            & 0.04 $\pm$ 0.01      & 0.01 $\pm$ 0.01      & 0.11 $\pm$ 0.03        & 0.09 $\pm$ 0.02        & 0.017 $\pm$ 0.003        \\ 
                      & $R^2$           & 0.877 $\pm$ 0.002       & -0.031 $\pm$ 0.0004      & 0.048 $\pm$ 0.003        & 0.387 $\pm$ 0.002        & 0.977 $\pm$ 0.001        \\ \hline

    \multirow{3}{*}{\makecell[t]{$P + B$ \\ 
    (\textbf{\mlp})}}                    
                      & RMSE            & 0.04 $\pm$  0.001 & 0.01 $\pm$  0.0001 & 0.095 $\pm$  0.002 & 0.078 $\pm$  0.002 & 0.029 $\pm$  0.001    \\ 
                      & $\chi^2$        & 1.42 $\pm$  0.14 & 1.02 $\pm$  0.05 & 1.19 $\pm$  0.09 & 1.26 $\pm$  0.15 & 0.91 $\pm$  0.14    \\ 
                      & $R^2$           & 0.89 $\pm$  0.01 & 0.24 $\pm$  0.02 & 0.31 $\pm$  0.03 & 0.55 $\pm$  0.02 & 0.93 $\pm$  0.01    \\ \hline

    \multirow{2}{*}{\makecell[t]{$P + B$ \\ 
    (\textbf{\gbt})}}
                    & RMSE            & 0.039 $\pm$ 0.009       & 0.01 $\pm$ 0.05       & 0.10 $\pm$ 0.04       & 0.08 $\pm$ 0.02        & 0.024 $\pm$ 0.002        \\ 
    
                      & $R^2$           & 0.897 $\pm$ 0.001       & 0.162 $\pm$ 0.001      & 0.223 $\pm$ 0.004        & 0.453 $\pm$ 0.003       & 0.952 $\pm$ 0.001        \\ \hline

    \multirow{3}{*}{\makecell[t]{Combined \\ 
    (\textbf{\hybrid})}}
                      & RMSE            & \textbf{0.023 $\pm$ 0.002} & 0.01 $\pm$ 0.001 & 0.096 $\pm$ 0.003 & \textbf{0.068 $\pm$ 0.005} & \textbf{0.011 $\pm$ 0.002}     \\ 
                      & $\chi^2$        & \textbf{1.47 $\pm$ 0.16} & 1.01 $\pm$ 0.11 & 1.15 $\pm$ 0.13 & \textbf{1.40 $\pm$ 0.24} & \textbf{0.80 $\pm$ 0.19}     \\ 
                      & $R^2$           & \textbf{0.97 $\pm$ 0.01} & 0.21 $\pm$ 0.10 & 0.30 $\pm$ 0.05 & \textbf{0.66 $\pm$ 0.05} & \textbf{0.99 $\pm$ 0.01}   \\ \hline         

    \multirow{2}{*}{\makecell[t]{Combined \\ 
    (\textbf{\gbt})}}
                      & RMSE            & 0.040 $\pm$ 0.001       & 0.011 $\pm$ 0.001      & 0.110 $\pm$ 0.004        & 0.093 $\pm$ 0.003        & 0.017 $\pm$ 0.001        \\   
                      & $R^2$           & 0.893 $\pm$ 0.006       & -0.024 $\pm$ 0.024     & 0.08 $\pm$ 0.06        & 0.36 $\pm$ 0.03        & 0.975 $\pm$ 0.002        \\ \hline
    \end{tabular}
    }
    \caption{Performance metrics for \cnn, \mlp, \hybrid (on their respective summary statistics) and \gbt (for the different statistics) for the LH dataset. Each value is the mean over 10 model initializations and the corresponding standard deviation. For boosted trees, the standard deviation comes from cross-validation with 4 different validation sets, keeping the training set size fixed.}
    \label{tab:main_metrics}
\end{table}

\paragraph{Results for standard cosmological parameters}

\begin{itemize}
    
\item Persistence images performed best for $\left\{\Omega_{\rm m}, \sigma_8\right\}$. In numbers, using PIs with the CNN achieved R$^2$ scores of $\left\{0.96,0.99\right\}$, compared to the \mlp's (using PS/BS) scores of $\left\{0.89,0.93\right\}$. On the other hand, none of the statistics are sensitive to $h$ and $\Omega_{\rm b}$, while both statistics achieve similar performance for $n_{\rm s}$. Looking at the $\chi^2$ values obtained, both \cnn and \mlp standard deviations seem to be well calibrated. 
    
\item 
When combining both summary statistics in the \hybrid architecture, we find that the RMSE on each parameter is close to that obtained using the data that best constrains it. This indicates that the power spectrum and bispectrum do not contain complementary information or not contained in the persistence images, as combining the three does not lead to significant improvement. 

\item Boosted trees lead to poorer performance than the neural networks, especially when involving persistence images (either alone or combined). We find that though they are cheaper to train, they tend to overfit in this setting, which impacts their generalization. This can be corrected by constraining the model flexibility, however at the cost of performance. Perhaps surprisingly, the neural-based methods seem to provide a better trade-off in this regime.
    
\item We claim that in our setup persistent homology in general does not probe scales beyond $2\pi/k_{\rm max}\approx21\,\text{Mpc}/h$,\footnote{A more detailed discussion can be found in \cite{Yip:2024hlz}, section 5.1.} the scale at which the power spectrum and bispectrum are truncated. While it is difficult to define a scale for a topological feature, we can use the birth value, which for the $1$- and $2$-cycles is half of the length of the edge that triggers the formation of the feature. Figure \ref{fig:pdpi} shows that most $1$- and $2$-cycles are born beyond $25\,\text{Mpc}/h$, and a topological feature should be of a size larger than this edge. For the $0$-cycles, the birth value corresponds inversely to the local halo number density, which we find to be within $20-25\,\text{Mpc}/h$, consistent with $\gtrsim21\,\text{Mpc}/h$, for the majority of the features. Hence, when persistent homology performs better than the power spectrum and bispectrum, we argue that the extra information extractable by the neural network model should not come from smaller-scale modes, but from the fact that persistent homology probes higher-order correlations given that each topological feature is typically formed from a set of many vertices.

\end{itemize}

\paragraph{PNG results}
\begin{itemize}
   \item Persistence images consistently perform better than the combination of power spectrum and bispectrum for retrieving the $\fnlloc$ parameter. The performance is significantly better when using \gbt, unlike the previous results on the LH. This observation requires more investigation to properly understand its underlying causes. This could be due to a combination of the variance differences between the two datasets (one varying 5 parameters, 3 of which have an effective impact on the summary statistics considered here, the other varying only a single parameter), the (relatively) small size of each dataset, the nature of the data and the properties of the methods.      
    \item All models struggled or failed to recover the $\fnleq$ value, and we therefore do not plot the Fiducial recovery. This seems to indicate that the summary statistics considered here are not sensitive enough to $\fnleq$ when marginalizing over the other parameters. Note that the $\fnleq$ inference case is much harder than the $\fnlloc$ case. The LH$\_\fnleq$ varies several cosmological parameters in addition to $\fnleq$, thus requiring to marginalize over those parameters. It would be interesting to investigate in future work whether integrating information about the other cosmological parameters (e.g., conditioning on the other parameters) helps in predicting $\fnleq$ with those summary statistics. 
    \item It is also important to note that $\fnleq$ is defined to give a primordial bispectrum of the same size as $\fnlloc$ at the equilateral configuration \cite{Babich:2004gb}. However, the local template is small for that configuration relative to other (more squeezed) configurations, so for equal values of $\fnlloc$ and $\fnleq$, local PNG has a larger physical effect. This effect is more pronounced for statistics that sum over many configurations, such as persistence images. 
    
    
    \item The other results on the LH\_$\fnleq$ dataset show a similar trend to the LH results regarding the informativeness of the persistence images. However, here we achieve best performance using the persistence images alone, with \cnn, for the parameters that can be constrained, such as $\left\{\Omega_{\rm m},\sigma_8,n_{\rm s}\right\}$. \gbt consistently underperform compared to the neural-based methods. Combining the PIs with the PS/BS in this setup leads to worse results. One possible explanation is that this setup enters a regime where the high dimensionality of the data, coupled with the smaller dataset size (half the size of the LH dataset), prevents the models from training properly.
    
\end{itemize}

\begin{table}[h!]
    \centering
    \begin{tabular}{|c|l| c|}
    \hline
    \textbf{Model} & Metric  & $\fnlloc$ \\[0.5ex]  \hline\hline

    \multirow{3}{*}{\makecell[t]{PI \\ 
    (\textbf{\cnn})}}
                   &  RMSE      &  47 $\pm$ 2     \\ 
                   &  $\chi^2$  & 1.48 $\pm$ 0.17  \\ 
                   &  $R^2$     &  0.93 $\pm$ 0.01 \\ \hline              
    \multirow{2}{*}{\makecell[t]{PI \\ 
    (\textbf{\gbt})}}
                   &  RMSE      &  38.3 $\pm$ 1.2\\  
                   &  $R^2$     & 0.950  $\pm$ 0.03\\ \hline

    \multirow{3}{*}{\makecell[t]{$P + B$ \\ 
    (\textbf{\mlp})}}
                   & RMSE        & 50 $\pm$ 2        \\ 
                  & $\chi^2$    & 2.40 $\pm$ 0.26    \\ 
                  &$R^2$        & 0.92 $\pm$ 0.01   \\ \hline

    \multirow{2}{*}{\makecell[t]{$P + B$ \\ 
    (\textbf{\gbt})}}                       
                  &  RMSE      &  48.8 $\pm$ 3.2 \\  
                   &  $R^2$     & 0.920 $\pm$ 0.01   \\ \hline

    \multirow{3}{*}{\makecell[t]{Combined \\ 
    (\textbf{\hybrid})}}
                  & RMSE        & 46 $\pm$ 2       \\ 
                  & $\chi^2$    & 1.58 $\pm$ 0.17    \\ 
                  & $R^2$       & 0.93 $\pm$ 0.01    \\ \hline
                  
    \multirow{2}{*}{\makecell[t]{Combined \\ 
    (\textbf{\gbt})}}               
                   &  RMSE      &  \textbf{37.6 $\pm$ 1.1}\\  
                   &  $R^2$     & \textbf{0.952 $\pm$ 0.003}   \\ \hline
    \end{tabular}
    \caption{Performance metrics for LH\_$\fnlloc$ on the test set for the different summary statistics and models. For neural networks, each value is the mean over 10 model initializations, with the standard deviation reported. For boosted trees, the standard deviation comes from 4-fold cross-validation.}
    \label{tab:fnl_loc_metrics}
\end{table}

\begin{table}[h!]
\hspace*{-1cm}
    \centering
    \small{
    \begin{tabular}{|c|l| c c c c c|}
    \hline
    \textbf{Model} & Metric   & $\fnleq$ & $\Omega_{\rm m}$  & $h$    & $n_{\rm s}$   & $\sigma_8$   \\[0.5ex]  \hline\hline
    \multirow{3}{*}{\makecell[t]{PI \\ 
    (\textbf{\cnn})}}
                   &  RMSE     & 340 $\pm$ 2 & \textbf{0.029 $\pm$ 0.005} & 0.11 $\pm$ 0.002 & \textbf{0.078 $\pm$ 0.005} & \textbf{0.018 $\pm$ 0.002}   \\ 
                   &  $\chi^2$ & 1.02 $\pm$ 0.02 & \textbf{1.02 $\pm$ 0.28} & 1.17 $\pm$ 0.10 & \textbf{1.19 $\pm$ 0.17} & \textbf{1.01 $\pm$ 0.23}   \\ 
                   &  $R^2$    & 0.005 $\pm$ 0.009 & \textbf{0.94 $\pm$ 0.03} & 0.22 $\pm$ 0.04 & \textbf{0.55 $\pm$ 0.06} &\textbf{0.98 $\pm$ 0.01}  \\ \hline

    \multirow{2}{*}{\makecell[t]{PI \\ 
    (\textbf{\gbt})}}
                      & RMSE            & 338 $\pm$ 10      & 0.050 $\pm$ 0.003      & 0.109 $\pm$ 0.004        & 0.096 $\pm$ 0.007        & 0.023 $\pm$ 0.002        \\ 
                      & $R^2$           & -0.01 $\pm$ 0.03       & 0.81 $\pm$ 0.03      & 0.13 $\pm$ 0.07        & 0.30 $\pm$ 0.05        & 0.963 $\pm$ 0.005        \\ \hline

    \multirow{3}{*}{\makecell[t]{$P + B$ \\ 
    (\textbf{\mlp})}}
                   & RMSE      & 340 $\pm$ 0.3 & 0.04 $\pm$ 0.001 & 0.089 $\pm$ 0.002 & 0.084 $\pm$ 0.003 & 0.035 $\pm$ 0.002    \\ 
                  & $\chi^2$   & 0.96 $\pm$ 0.01 & 0.98 $\pm$ 0.07 & 1.36 $\pm$ 0.08 & 1.14 $\pm$ 0.07 & 1.11 $\pm$ 0.12   \\ 
                  &$R^2$       & 0.002 $\pm$ 0.002 & 0.88 $\pm$ 0.01 & 0.42 $\pm$ 0.02 & 0.47 $\pm$ 0.04 & 0.91 $\pm$ 0.01 \\ \hline

    \multirow{2}{*}{\makecell[t]{$P + B$ \\ 
    (\textbf{\gbt})}}
                    & RMSE            & 333 $\pm$ 14       & 0.042 $\pm$ 0.001       & 0.104 $\pm$ 0.003       & 0.081 $\pm$ 0.006        & 0.028 $\pm$ 0.004        \\ 
                    & $R^2$           & 0.02 $\pm$ 0.02       & 0.86 $\pm$ 0.01      & 0.21 $\pm$ 0.06        & 0.50 $\pm$ 0.06       & 0.94 $\pm$ 0.01        \\ \hline

    \multirow{3}{*}{\makecell[t]{Combined \\ 
    (\textbf{\hybrid})}}
                  & RMSE       & 340 $\pm$ 0.3 & 0.04 $\pm$ 0.001 & 0.089 $\pm$ 0.002 & 0.084 $\pm$ 0.003 & 0.035 $\pm$ 0.002    \\ 
                  & $\chi^2$   & 0.96 $\pm$ 0.01 & 0.98 $\pm$ 0.07 & 1.36 $\pm$ 0.08 & 1.14 $\pm$ 0.07 & 1.11 $\pm$ 0.12   \\ 
                  & $R^2$      &0.002 $\pm$ 0.002 & 0.88 $\pm$ 0.01 & 0.42 $\pm$ 0.02 & 0.47 $\pm$ 0.04 & 0.91 $\pm$ 0.01   \\ \hline

    \multirow{2}{*}{\makecell[t]{Combined \\ 
    (\textbf{\gbt})}}
                      & RMSE            & 338 $\pm$ 10       & 0.046 $\pm$ 0.003      & 0.109 $\pm$ 0.004        & 0.092 $\pm$ 0.007        & 0.023 $\pm$ 0.006        \\   
                      & $R^2$           & -0.01 $\pm$ 0.04       & 0.84 $\pm$ 0.02     & 0.13 $\pm$ 0.07        & 0.35 $\pm$ 0.06        & 0.962 $\pm$ 0.005        \\ \hline

    \end{tabular}}
    \caption{Performance Metrics Comparison for LH\_$\fnleq$ on the test set, for the different summary statistics and models. For neural networks, each value is the mean over 10 model initializations and the standard deviation over the mean. For boosted trees, the standard deviation comes from 4-fold cross-validation.}
   \label{tab:fnl_eq_metrics}
\end{table}

\subsection{Fiducial parameter recovery}

\begin{figure}[]
    \centering
    \includegraphics[width=1\textwidth]{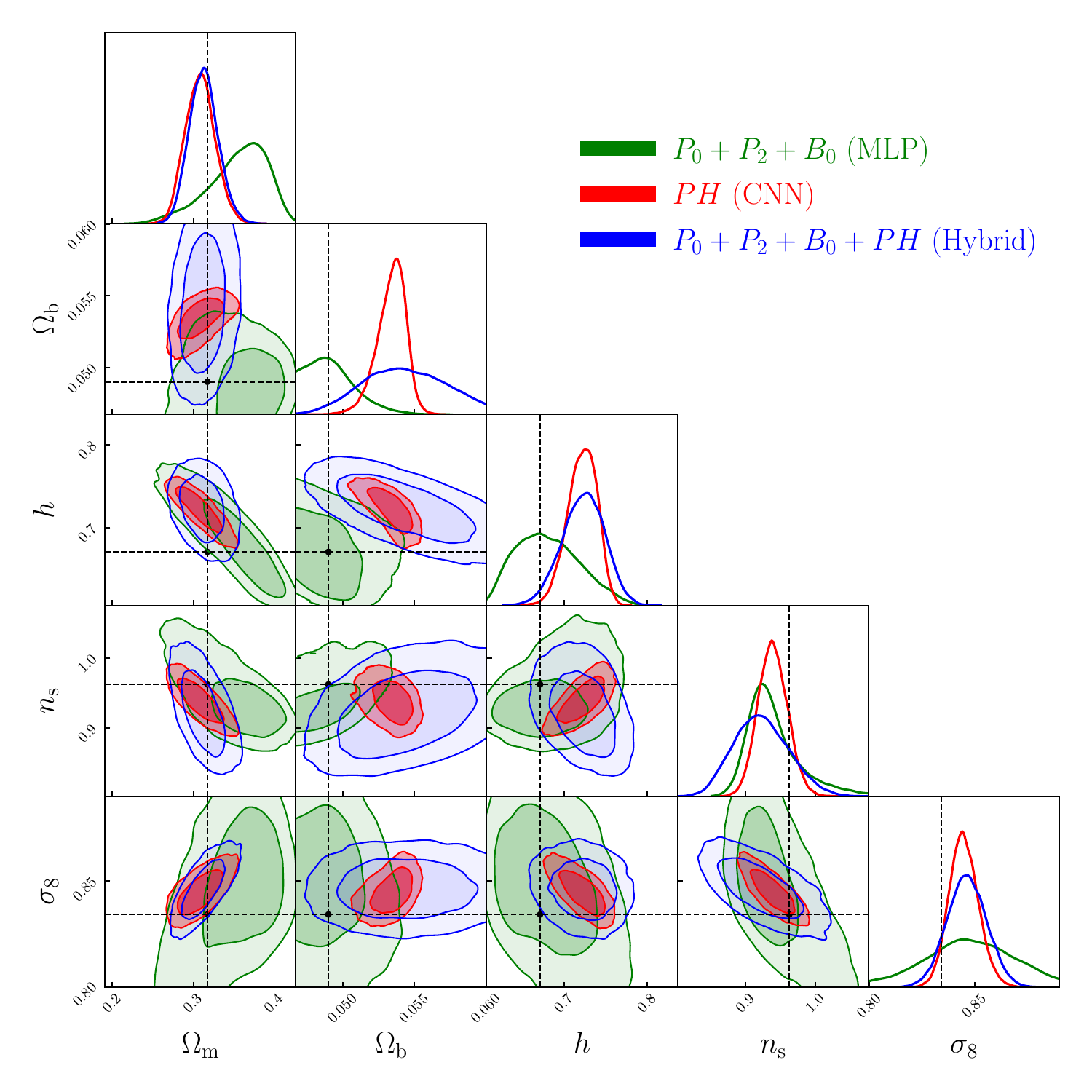} 
    \caption{Predictions of cosmological parameters on the Fiducial dataset using the best-performing models trained on the \quijote Latin Hypercube (LH) at redshift $z=0.5$. The contours indicate the regions containing $68\%$ ($1$-$\sigma$) and $95\%$ ($2$-$\sigma$) of the predictions. The crosshairs indicate the true fiducial values. The diagonal panels show the $1$D distributions.}
    \label{fig:main_fiducial_recovery}
\end{figure}

\begin{figure}[]
    \centering
    \includegraphics[width=0.5\textwidth]{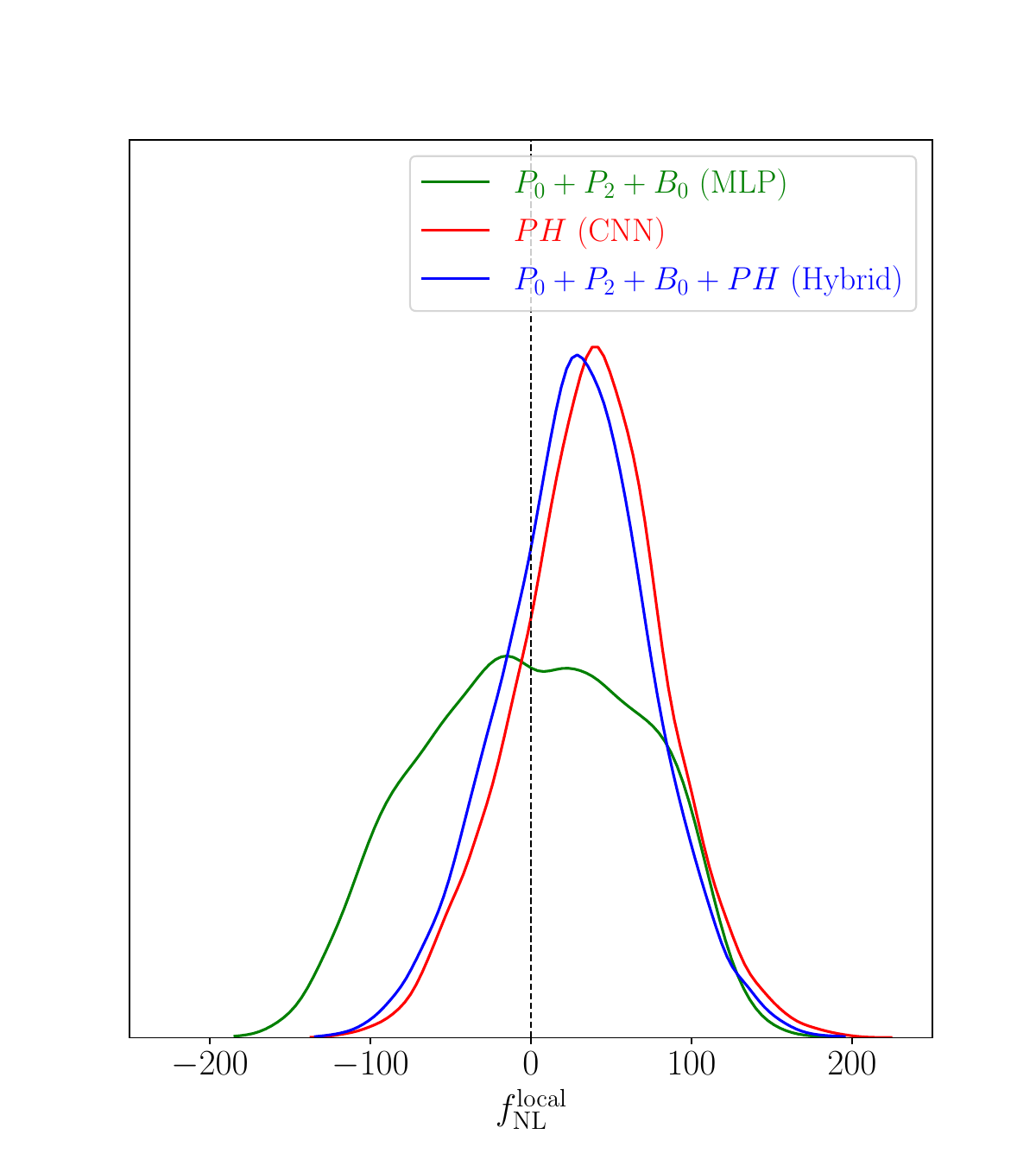} 
    \caption{Distribution of $\fnlloc$ predictions using the best-performing models trained on the \quijotepng Latin Hypercube (LH\_$\fnlloc$) at redshift $z=0.5$. The dashed line marks the fiducial value $\fnlloc=0$.}
    \label{fig:fiducial_recovery_loc}
\end{figure}

We now turn our attention to the fiducial dataset. We want to evaluate how well our models constrain parameters at the fiducial cosmology. The dataset contains $15{,}000$ fiducial realizations never seen by the models, and from them we generate the same number of predictions for each best-performing model. The predictions are plotted in Figures \ref{fig:main_fiducial_recovery} and \ref{fig:fiducial_recovery_loc}, for models trained on LH and LH\_$\fnlloc$ respectively (we do not plot for the models trained on LH\_$\fnleq$ as they fail to infer $\fnleq$ completely), as contours marking the $68\%$ and $95\%$ confidence regions, as well as each parameter's $1$D distribution. We report the following results:

\begin{itemize}
    \item Persistence images outperform power spectrum and bispectrum combined in both accuracy and precision for \{$\Omega_{\rm m}$, $n_{\rm s}$, $\sigma_8$\}. This is expected since the model achieves better $R^2$ values for those parameters.
    \item For \{$\Omega_{\rm b}$, $h$\}, predictions from the power spectrum and bispectrum combined are well centered on the fiducial value; however, this should not be overinterpreted, as the models have poor $R^2$ scores overall for these parameters. The model simply learns better at using the mean of the parameter values it trains on as the prediction, which happens to be the fiducial value.
    \item The \hybrid model generally does not improve on either of the other models that use a single type of statistics, except for \{$\Omega_{\rm m}$, $\fnlloc$\}.
    \item The contours seem offset with respect to the fiducial values for parameters which are not well constrained. This does not necessarily mean that the model is biased since the contours do not include information about the estimated variance on the prediction.
\end{itemize}

\subsection{Understanding persistence images with feature importance}
\label{sec:featuresimportance}
Features importance is a potentially useful tool for understanding how machine learning models link input data to the outcomes. From tree-based models, it is possible to extract an importance score for each input feature, quantifying its contribution to the model's performance.  



\begin{figure}[]
    \centering
    \caption*{Persistence Image (PI) - LH\_$\fnlloc$ dataset - \gbt}
    \includegraphics[width=1\textwidth]{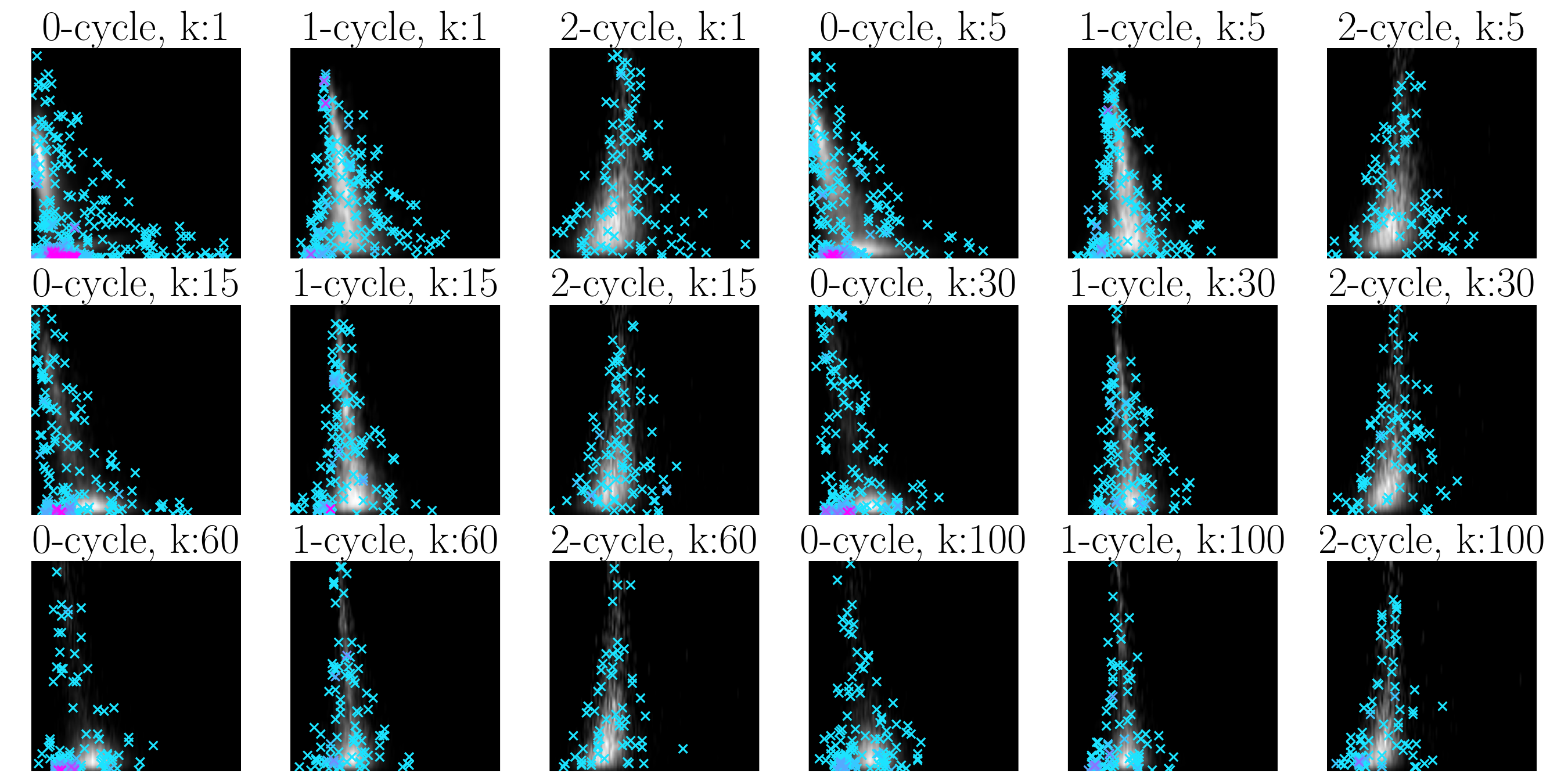}

    \caption{Feature importance map for the \gbt model trained on the LH\_$\fnlloc$ dataset to predict $\fnlloc$. Crosses indicate the positions of non-null features, with cyan marking pixels where the feature weight is non-null and below 1, and pink denoting weights above 1. The background shows the persistence image for a realization with $\fnlloc = -288.3$, providing context for the highlighted features. Each image represents a different input channel, corresponding to all cycles with varying $k-$neighbors in the filtration.}
    \label{fig:feature_map}
\end{figure}

\begin{figure}[]
    \centering
    \caption*{Persistence Image (PI)  - LH dataset - \gbt}
    \includegraphics[width=1\textwidth]{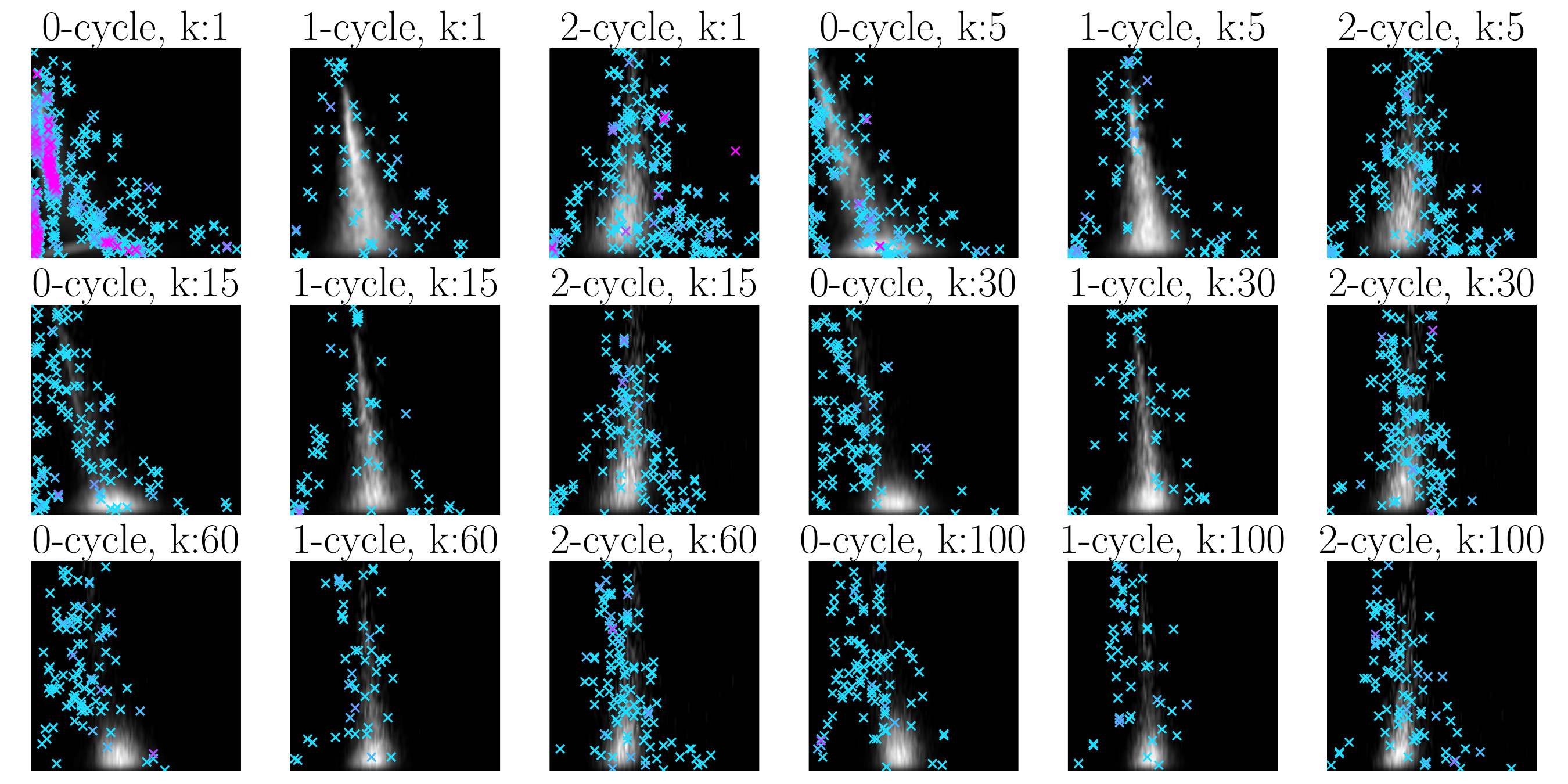}

    \caption{Feature importance map for the \gbt model trained on the LH dataset, focused exclusively on $\Omega_{\rm m}$.  Crosses indicate the positions of non-null features, with cyan marking pixels where the feature weight is non-null and below 1 and pink denoting weights above 1. The background shows the persistence image for a realization with $\Omega_{\rm m} = 0.3227$. Each image represents a different input channel, corresponding to all cycles with varying $k-$neighbors in the filtration.}
    \label{fig:feature_map_omegam}
\end{figure}

The feature importance map generated by the \gbt model identifies the pixels with the greatest influence on the model's predictions. In \xgboost, the ``weight'' of a feature refers to the number of times that feature is selected for a split across all decision trees in the ensemble, thus essentially measuring how frequently a feature contributes to the decision-making process within the model. Each tree participating in the ensemble only uses a few features, such that only a few pixels are assigned a non-zero weight. This score is computed directly by the \xgboost package. In Fig.~\ref{fig:feature_map}, 
we plot a map for each of the 3 input cycles and for each of the $k$---which determines the number of neighbors considered in the filtration, as described earlier. For visualization purposes, the background shows the corresponding persistence image for a realization with $\fnlloc = -288.3$. Crosses mark the positions of important features for predicting $\fnlloc$, with the color indicating their importance score: cyan crosses correspond to non-null pixels with a weight below 1, whereas pink crosses represent pixels with a weight above 1. Figure \ref{fig:feature_map_omegam} shows a similar plot for a model trained to predict $\Omega_{\rm m}$ only. 

This visualization shows that regions associated with early-birth, low-persistence features in 0-cycles, as well as the borders of the persistence image, are critical for the model's predictions. For example, these areas exhibit the largest variations in the $\fnlloc$ parameter (see Figure 8 in \cite{Yip:2024hlz}). These regions emphasize information tied to overdense regions, particularly those associated with early-forming structures. These features, identified early in the persistence pipeline, are often associated with ``independent'' massive halos. We see that the model trained on $\fnlloc$ uses a relatively greater number of features from 1-cycles compared to $\Omega_{\rm m}$. Most of the information on $\Omega_{\rm m}$ seems to be contained in the 0- and 2-cycles (which roughly correspond to clusters and voids respectively).

This is a first attempt at understanding the persistence image summary statistic. It would be interesting to explore the robustness of these conclusions with a larger dataset.

\section{Conclusions and outlook}
\label{sec:conclusions}

This paper investigates the potential of persistence images (PIs), along with their combination with power spectrum and bispectrum (PS/BS) data, for cosmological parameter inference. We trained and compared these summary statistics using different types of machine learning methods (neural-based and tree-based) to assess their strengths and limitations in extracting fundamental cosmological parameters and primordial non-Gaussianity amplitudes. 

Our results demonstrate that persistence images are highly effective for recovering cosmological parameters under various scenarios. The information contained in persistence images consistently outperformed other methods for parameters such as $\Omega_{\rm m}$ and $\sigma_8$, achieving higher $R^2$ values and lower RMSE compared to the PS/BS data. Notably, persistence images were particularly effective in constraining $\fnlloc$. Interestingly, \gbt provided a computationally cheaper alternative and, for $\fnlloc$, outperformed neural networks when trained on the same dataset.

As highlighted in \cite{bermejo2024topologicalbiashaloestrace}, persistent homology reveals clustering differences that are not captured by second-order statistics or local density alone, but instead reflect the higher-order connectivity of the cosmic web. Because cosmological models imprint subtle changes not only in how much structure forms but also in how that structure connects and surrounds matter, topological measures can detect morphological differences that traditional summary statistics often compress, making them potentially more sensitive and complementary for cosmological parameter inference. This is particularly relevant for parameters such as $\Omega_{\rm m}$, $\sigma_8$, and $\fnlloc$, which physically influence both the halo distribution and the surrounding matter field.

Combining PI and PS/BS data in the \hybrid model resulted in only marginal improvements over the \cnn alone. This suggests that the PS/BS do not contain additional information relevant to parameter inference beyond what is already present in the PIs, nor do they offer complementary information. However, our analysis is limited by the properties of the available datasets: they are relatively small from a machine learning standpoint, which forces us to limit the flexibility of the models considered to prevent overfitting. The different LHs considered here also have different properties in terms of number of parameters varied. 
It would be interesting to explore the robustness of the PI summary statistic to marginalization over other cosmological parameters.



Despite these successes, all models struggled to constrain $\fnleq$. This may be because the LH\_$\fnleq$ dataset varies several cosmological parameters over a large range. It would also be interesting to explore whether this statistic contains information about $\fnleq$ by including tighter priors on cosmological parameters such as $\Omega_{\rm m}$.

Parameters such as $\fnlloc$, $\Omega_{\rm m}$, and $\sigma_8$ tend to increase the number of halos and amplify clustering features that are more readily captured by persistence images.
In contrast, $\fnleq$ has a subtler effect, and improving its constraints would require larger simulation volumes to reduce sample variance. Notably, persistent homology offers a distinct advantage by focusing on small-scale features that remain sensitive to equilateral non-Gaussianity without relying on large-scale modes. Increasing simulation volumes could significantly enhance constraints on $\fnleq$ by enabling the detection of more features across multiple filtration scales. It was argued in \cite{Biagetti:2022qjl} that this can be achieved by joining many small simulations since PIs don't necessarily derive their constraining power from the large scales.

While direct comparisons to other studies are complicated by differences in dataset preparation --such as variations in power spectrum/bispectrum measurements, redshift ranges, mass cuts, and Fourier binning-- our findings remain broadly consistent with related work  \cite{Valogiannis:2021chp,Chatterjee:2024eur,Massara:2024cvu,Jung:2024esv,Jung:2023kjh,Peron:2024xaw,Makinen:2022jsc,Hortua:2023kuw,Ho:2024whi,Cuesta-Lazaro:2023zuk,Min:2024dgd}.

As shown in \cite{bermejo2024topologicalbiashaloestrace}, persistent homology is sensitive to the halo mass range, with more massive halos typically tracing larger-scale topological features than their lower-mass counterparts. Future work should explicitly evaluate how variations in halo populations influence the constraining power of topological summary statistics. Moreover, our feature importance analyses with \xgboost indicate that the model primarily focuses on early-born topological features, those that emerge early in the filtration process. In our setup, these early features correspond to haloes embedded in high-density regions, which are often associated with massive haloes. This suggests that the model may be capturing parameter dependencies through mass-related topological imprints. Given that cosmological parameters such as $\Omega_{\rm m}$ and $\fnlloc$ strongly affect the abundance of high-mass haloes, we anticipate that different halo populations will vary in their constraining power, with some providing stronger sensitivity to specific cosmological parameters than others.




Direct inference using persistence diagrams through methods such as \texttt{DeepSets} or \texttt{PersLay} offers a promising direction for future research \cite{zaheer2017deep,carriere2020perslay}. While substantial performance gains are not guaranteed, these approaches could reveal complementary information and improve our understanding of how persistence features encode cosmological information.

Transitioning to simulation-based inference frameworks, such as those outlined in recent studies \cite{cranmer2020frontier,Reza:2024djq,Miller:2021hys,Massara:2024cvu,Mootoovaloo:2024sao,Makinen:2024xph,khullar2022digs,Lehman:2024vyl,Saxena:2024rhu,Hahn:2022zxa,Tucci:2023bag}, could further enhance parameter estimation by enabling direct sampling of posterior distributions. Successfully integrating these methods with convolutional neural networks or tree-based models like \xgboost will require careful development and experimentation, but they hold significant potential for advancing the field of cosmological inference.

\acknowledgments
We thank Moritz Münchmeyer for useful discussions and access to his group's computational resources in the early stages of this project. We thank Matteo Biagetti, Mathieu Carri\`ere, and Francisco Villaescusa-Navarro for useful discussions and feedback on this project. The work of J.H.T.Y. and G.S. is supported by the U.S. Department of Energy, Office of Science, Office of High Energy Physics under Award Numbers DE-SC-0023719 and DE-SC-0017647. J.N. is supported by FONDECYT Regular grant 1211545. J.C. is supported by FONDECYT de Postdoctorado, N° 3240444. Powered@NLHPC: This research was partially supported by the supercomputing infrastructure of NLHPC (CCSS210001). G.C. acknowledges support from the European Union's Horizon Europe research and innovation programme under the Marie Skłodowska-Curie Postdoctoral Fellowship Programme, SMASH co-funded under the grant agreement No. 101081355.

We ran test simulations, all persistent homology calculations, and power spectrum and bispectrum measurements using the computing resources and assistance of the University of Wisconsin-Madison Center for High Throughput Computing (CHTC) \cite{https://doi.org/10.21231/gnt1-hw21} in the Department of Computer Sciences. The CHTC is supported by UW-Madison, the Advanced Computing Initiative, the Wisconsin Alumni Research Foundation, the Wisconsin Institutes for Discovery, and the National Science Foundation, and is an active member of the OSG Consortium, which is supported by the National Science Foundation and the U.S. Department of Energy's Office of Science.


\bibliographystyle{JHEP}
\bibliography{biblio.bib}

\providecommand{\href}[2]{#2}\begingroup\raggedright\begin{thebibliography}{100}

\bibitem{Yip:2023vud}
J.H.T.~Yip, A.~Rouhiainen and G.~Shiu, \emph{{Learning from Topology: Cosmological Parameter Estimation from the Large-scale Structure}},  in \emph{{40th International Conference on Machine Learning}}, 8, 2023 [\href{https://arxiv.org/abs/2308.02636}{{\ttfamily 2308.02636}}].

\bibitem{SPHEREx:2014bgr}
{\scshape SPHEREx} collaboration, \emph{{Cosmology with the SPHEREX All-Sky Spectral Survey}},  \href{https://arxiv.org/abs/1412.4872}{{\ttfamily 1412.4872}}.

\bibitem{Amendola:2016saw}
L.~Amendola et~al., \emph{{Cosmology and fundamental physics with the Euclid satellite}}, \href{https://doi.org/10.1007/s41114-017-0010-3}{\emph{Living Rev. Rel.} {\bfseries 21} (2018) 2} [\href{https://arxiv.org/abs/1606.00180}{{\ttfamily 1606.00180}}].

\bibitem{Zhan:2017uwu}
H.~Zhan and J.A.~Tyson, \emph{{Cosmology with the Large Synoptic Survey Telescope: an Overview}}, \href{https://doi.org/10.1088/1361-6633/aab1bd}{\emph{Rept. Prog. Phys.} {\bfseries 81} (2018) 066901} [\href{https://arxiv.org/abs/1707.06948}{{\ttfamily 1707.06948}}].

\bibitem{Ivanov:2019pdj}
M.M.~Ivanov, M.~Simonovi\'c and M.~Zaldarriaga, \emph{{Cosmological Parameters from the BOSS Galaxy Power Spectrum}}, \href{https://doi.org/10.1088/1475-7516/2020/05/042}{\emph{JCAP} {\bfseries 05} (2020) 042} [\href{https://arxiv.org/abs/1909.05277}{{\ttfamily 1909.05277}}].

\bibitem{Zhang:2021yna}
P.~Zhang, G.~D'Amico, L.~Senatore, C.~Zhao and Y.~Cai, \emph{{BOSS Correlation Function analysis from the Effective Field Theory of Large-Scale Structure}}, \href{https://doi.org/10.1088/1475-7516/2022/02/036}{\emph{JCAP} {\bfseries 02} (2022) 036} [\href{https://arxiv.org/abs/2110.07539}{{\ttfamily 2110.07539}}].

\bibitem{Ivanov:2019hqk}
M.M.~Ivanov, M.~Simonovi\'c and M.~Zaldarriaga, \emph{{Cosmological Parameters and Neutrino Masses from the Final Planck and Full-Shape BOSS Data}}, \href{https://doi.org/10.1103/PhysRevD.101.083504}{\emph{Phys. Rev. D} {\bfseries 101} (2020) 083504} [\href{https://arxiv.org/abs/1912.08208}{{\ttfamily 1912.08208}}].

\bibitem{DES:2020ahh}
{\scshape DES} collaboration, \emph{{Dark Energy Survey Year 1 Results: Cosmological constraints from cluster abundances and weak lensing}}, \href{https://doi.org/10.1103/PhysRevD.102.023509}{\emph{Phys. Rev. D} {\bfseries 102} (2020) 023509} [\href{https://arxiv.org/abs/2002.11124}{{\ttfamily 2002.11124}}].

\bibitem{DES:2020mlx}
{\scshape DES} collaboration, \emph{{Dark Energy Survey Year 1 Results: Cosmological Constraints from Cluster Abundances, Weak Lensing, and Galaxy Correlations}}, \href{https://doi.org/10.1103/PhysRevLett.126.141301}{\emph{Phys. Rev. Lett.} {\bfseries 126} (2021) 141301} [\href{https://arxiv.org/abs/2010.01138}{{\ttfamily 2010.01138}}].

\bibitem{Vikhlinin:2008ym}
A.~Vikhlinin et~al., \emph{{Chandra Cluster Cosmology Project III: Cosmological Parameter Constraints}}, \href{https://doi.org/10.1088/0004-637X/692/2/1060}{\emph{Astrophys. J.} {\bfseries 692} (2009) 1060} [\href{https://arxiv.org/abs/0812.2720}{{\ttfamily 0812.2720}}].

\bibitem{rozo2009cosmological}
E.~Rozo, R.H.~Wechsler, E.S.~Rykoff, J.T.~Annis, M.R.~Becker, A.E.~Evrard et~al., \emph{Cosmological constraints from the sloan digital sky survey maxbcg cluster catalog}, {\emph{The Astrophysical Journal} {\bfseries 708} (2009) 645}.

\bibitem{SPT:2024qbr}
{\scshape SPT, DES} collaboration, \emph{{SPT clusters with DES and HST weak lensing. II. Cosmological constraints from the abundance of massive halos}}, \href{https://doi.org/10.1103/PhysRevD.110.083510}{\emph{Phys. Rev. D} {\bfseries 110} (2024) 083510} [\href{https://arxiv.org/abs/2401.02075}{{\ttfamily 2401.02075}}].

\bibitem{Ghirardini:2024yni}
V.~Ghirardini et~al., \emph{{The SRG/eROSITA all-sky survey - Cosmology constraints from cluster abundances in the western Galactic hemisphere}}, \href{https://doi.org/10.1051/0004-6361/202348852}{\emph{Astron. Astrophys.} {\bfseries 689} (2024) A298} [\href{https://arxiv.org/abs/2402.08458}{{\ttfamily 2402.08458}}].

\bibitem{eBOSS:2020yzd}
{\scshape eBOSS} collaboration, \emph{{Completed SDSS-IV extended Baryon Oscillation Spectroscopic Survey: Cosmological implications from two decades of spectroscopic surveys at the Apache Point Observatory}}, \href{https://doi.org/10.1103/PhysRevD.103.083533}{\emph{Phys. Rev. D} {\bfseries 103} (2021) 083533} [\href{https://arxiv.org/abs/2007.08991}{{\ttfamily 2007.08991}}].

\bibitem{Verde:2001sf}
L.~Verde et~al., \emph{{The 2dF Galaxy Redshift Survey: The Bias of galaxies and the density of the Universe}}, \href{https://doi.org/10.1046/j.1365-8711.2002.05620.x}{\emph{Mon. Not. Roy. Astron. Soc.} {\bfseries 335} (2002) 432} [\href{https://arxiv.org/abs/astro-ph/0112161}{{\ttfamily astro-ph/0112161}}].

\bibitem{Gil-Marin:2014sta}
H.~Gil-Mar\'\i{}n, J.~Nore\~na, L.~Verde, W.J.~Percival, C.~Wagner, M.~Manera et~al., \emph{{The power spectrum and bispectrum of SDSS DR11 BOSS galaxies \textendash{} I. Bias and gravity}}, \href{https://doi.org/10.1093/mnras/stv961}{\emph{Mon. Not. Roy. Astron. Soc.} {\bfseries 451} (2015) 539} [\href{https://arxiv.org/abs/1407.5668}{{\ttfamily 1407.5668}}].

\bibitem{DAmico:2019fhj}
G.~D'Amico, J.~Gleyzes, N.~Kokron, K.~Markovic, L.~Senatore, P.~Zhang et~al., \emph{{The Cosmological Analysis of the SDSS/BOSS data from the Effective Field Theory of Large-Scale Structure}}, \href{https://doi.org/10.1088/1475-7516/2020/05/005}{\emph{JCAP} {\bfseries 05} (2020) 005} [\href{https://arxiv.org/abs/1909.05271}{{\ttfamily 1909.05271}}].

\bibitem{DAmico:2022osl}
G.~D'Amico, Y.~Donath, M.~Lewandowski, L.~Senatore and P.~Zhang, \emph{{The BOSS bispectrum analysis at one loop from the Effective Field Theory of Large-Scale Structure}}, \href{https://doi.org/10.1088/1475-7516/2024/05/059}{\emph{JCAP} {\bfseries 05} (2024) 059} [\href{https://arxiv.org/abs/2206.08327}{{\ttfamily 2206.08327}}].

\bibitem{Ivanov:2023qzb}
M.M.~Ivanov, O.H.E.~Philcox, G.~Cabass, T.~Nishimichi, M.~Simonovi\'c and M.~Zaldarriaga, \emph{{Cosmology with the galaxy bispectrum multipoles: Optimal estimation and application to BOSS data}}, \href{https://doi.org/10.1103/PhysRevD.107.083515}{\emph{Phys. Rev. D} {\bfseries 107} (2023) 083515} [\href{https://arxiv.org/abs/2302.04414}{{\ttfamily 2302.04414}}].

\bibitem{Colas:2019ret}
T.~Colas, G.~D'amico, L.~Senatore, P.~Zhang and F.~Beutler, \emph{{Efficient Cosmological Analysis of the SDSS/BOSS data from the Effective Field Theory of Large-Scale Structure}}, \href{https://doi.org/10.1088/1475-7516/2020/06/001}{\emph{JCAP} {\bfseries 06} (2020) 001} [\href{https://arxiv.org/abs/1909.07951}{{\ttfamily 1909.07951}}].

\bibitem{Biagetti:2019bnp}
M.~Biagetti, \emph{{The Hunt for Primordial Interactions in the Large Scale Structures of the Universe}}, \href{https://doi.org/10.3390/galaxies7030071}{\emph{Galaxies} {\bfseries 7} (2019) 71} [\href{https://arxiv.org/abs/1906.12244}{{\ttfamily 1906.12244}}].

\bibitem{Cabass:2022ymb}
G.~Cabass, M.M.~Ivanov, O.H.E.~Philcox, M.~Simonovi\'c and M.~Zaldarriaga, \emph{{Constraints on multifield inflation from the BOSS galaxy survey}}, \href{https://doi.org/10.1103/PhysRevD.106.043506}{\emph{Phys. Rev. D} {\bfseries 106} (2022) 043506} [\href{https://arxiv.org/abs/2204.01781}{{\ttfamily 2204.01781}}].

\bibitem{Cabass:2022wjy}
G.~Cabass, M.M.~Ivanov, O.H.E.~Philcox, M.~Simonovi\'c and M.~Zaldarriaga, \emph{{Constraints on Single-Field Inflation from the BOSS Galaxy Survey}},  \href{https://arxiv.org/abs/2201.07238}{{\ttfamily 2201.07238}}.

\bibitem{DAmico:2022gki}
G.~D'Amico, M.~Lewandowski, L.~Senatore and P.~Zhang, \emph{{Limits on primordial non-Gaussianities from BOSS galaxy-clustering data}},  \href{https://arxiv.org/abs/2201.11518}{{\ttfamily 2201.11518}}.

\bibitem{White:2016yhs}
M.~White, \emph{{A marked correlation function for constraining modified gravity models}}, \href{https://doi.org/10.1088/1475-7516/2016/11/057}{\emph{JCAP} {\bfseries 11} (2016) 057} [\href{https://arxiv.org/abs/1609.08632}{{\ttfamily 1609.08632}}].

\bibitem{Jung:2024esv}
G.~Jung, A.~Ravenni, M.~Liguori, M.~Baldi, W.R.~Coulton, F.~Villaescusa-Navarro et~al., \emph{{Quijote-PNG: Optimizing the summary statistics to measure Primordial non-Gaussianity}},  \href{https://arxiv.org/abs/2403.00490}{{\ttfamily 2403.00490}}.

\bibitem{Massara:2020pli}
E.~Massara, F.~Villaescusa-Navarro, S.~Ho, N.~Dalal and D.N.~Spergel, \emph{{Using the Marked Power Spectrum to Detect the Signature of Neutrinos in Large-Scale Structure}}, \href{https://doi.org/10.1103/PhysRevLett.126.011301}{\emph{Phys. Rev. Lett.} {\bfseries 126} (2021) 011301} [\href{https://arxiv.org/abs/2001.11024}{{\ttfamily 2001.11024}}].

\bibitem{Marinucci:2024bdq}
M.~Marinucci et~al., \emph{{The constraining power of the Marked Power Spectrum: an analytical study}},  \href{https://arxiv.org/abs/2411.14377}{{\ttfamily 2411.14377}}.

\bibitem{Massara:2022zrf}
E.~Massara, F.~Villaescusa-Navarro, C.~Hahn, M.M.~Abidi, M.~Eickenberg, S.~Ho et~al., \emph{{Cosmological Information in the Marked Power Spectrum of the Galaxy Field}}, \href{https://doi.org/10.3847/1538-4357/acd44d}{\emph{Astrophys. J.} {\bfseries 951} (2023) 70} [\href{https://arxiv.org/abs/2206.01709}{{\ttfamily 2206.01709}}].

\bibitem{Cowell:2024wyl}
J.A.~Cowell, D.~Alonso and J.~Liu, \emph{{Optimizing marked power spectra for cosmology}}, \href{https://doi.org/10.1093/mnras/stae2492}{\emph{Mon. Not. Roy. Astron. Soc.} {\bfseries 535} (2024) 3129} [\href{https://arxiv.org/abs/2409.05695}{{\ttfamily 2409.05695}}].

\bibitem{Hou:2024blc}
J.~Hou, A.~Moradinezhad~Dizgah, C.~Hahn, M.~Eickenberg, S.~Ho, P.~Lemos et~al., \emph{{Cosmological constraints from the redshift-space galaxy skew spectra}}, \href{https://doi.org/10.1103/PhysRevD.109.103528}{\emph{Phys. Rev. D} {\bfseries 109} (2024) 103528} [\href{https://arxiv.org/abs/2401.15074}{{\ttfamily 2401.15074}}].

\bibitem{Schmittfull:2020hoi}
M.~Schmittfull and A.~Moradinezhad~Dizgah, \emph{{Galaxy skew-spectra in redshift-space}}, \href{https://doi.org/10.1088/1475-7516/2021/03/020}{\emph{JCAP} {\bfseries 03} (2021) 020} [\href{https://arxiv.org/abs/2010.14267}{{\ttfamily 2010.14267}}].

\bibitem{Peron:2024xaw}
M.~Peron, G.~Jung, M.~Liguori and M.~Pietroni, \emph{{Constraining primordial non-Gaussianity from large scale structure with the wavelet scattering transform}}, \href{https://doi.org/10.1088/1475-7516/2024/07/021}{\emph{JCAP} {\bfseries 07} (2024) 021} [\href{https://arxiv.org/abs/2403.17657}{{\ttfamily 2403.17657}}].

\bibitem{Eickenberg:2022qvy}
M.~Eickenberg et~al., \emph{{Wavelet Moments for Cosmological Parameter Estimation}},  \href{https://arxiv.org/abs/2204.07646}{{\ttfamily 2204.07646}}.

\bibitem{Valogiannis:2022xwu}
G.~Valogiannis and C.~Dvorkin, \emph{{Going beyond the galaxy power spectrum: An analysis of BOSS data with wavelet scattering transforms}}, \href{https://doi.org/10.1103/PhysRevD.106.103509}{\emph{Phys. Rev. D} {\bfseries 106} (2022) 103509} [\href{https://arxiv.org/abs/2204.13717}{{\ttfamily 2204.13717}}].

\bibitem{Valogiannis:2023mxf}
G.~Valogiannis, S.~Yuan and C.~Dvorkin, \emph{{Precise cosmological constraints from BOSS galaxy clustering with a simulation-based emulator of the wavelet scattering transform}}, \href{https://doi.org/10.1103/PhysRevD.109.103503}{\emph{Phys. Rev. D} {\bfseries 109} (2024) 103503} [\href{https://arxiv.org/abs/2310.16116}{{\ttfamily 2310.16116}}].

\bibitem{Uhlemann:2019gni}
C.~Uhlemann, O.~Friedrich, F.~Villaescusa-Navarro, A.~Banerjee and S.~Codis, \emph{{Fisher for complements: Extracting cosmology and neutrino mass from the counts-in-cells PDF}}, \href{https://doi.org/10.1093/mnras/staa1155}{\emph{Mon. Not. Roy. Astron. Soc.} {\bfseries 495} (2020) 4006} [\href{https://arxiv.org/abs/1911.11158}{{\ttfamily 1911.11158}}].

\bibitem{Friedrich:2019byw}
O.~Friedrich, C.~Uhlemann, F.~Villaescusa-Navarro, T.~Baldauf, M.~Manera and T.~Nishimichi, \emph{{Primordial non-Gaussianity without tails \textendash{} how to measure fNL with the bulk of the density PDF}}, \href{https://doi.org/10.1093/mnras/staa2160}{\emph{Mon. Not. Roy. Astron. Soc.} {\bfseries 498} (2020) 464} [\href{https://arxiv.org/abs/1912.06621}{{\ttfamily 1912.06621}}].

\bibitem{Gould:2024sve}
B.M.~Gould, L.~Castiblanco, C.~Uhlemann and O.~Friedrich, \emph{{Cosmology on point: modelling spectroscopic tracer one-point statistics}},  \href{https://arxiv.org/abs/2409.18182}{{\ttfamily 2409.18182}}.

\bibitem{DAmico:2010dwy}
G.~D'Amico, M.~Musso, J.~Norena and A.~Paranjape, \emph{{Excursion Sets and Non-Gaussian Void Statistics}}, \href{https://doi.org/10.1103/PhysRevD.83.023521}{\emph{Phys. Rev. D} {\bfseries 83} (2011) 023521} [\href{https://arxiv.org/abs/1011.1229}{{\ttfamily 1011.1229}}].

\bibitem{Kamionkowski:2008sr}
M.~Kamionkowski, L.~Verde and R.~Jimenez, \emph{{The Void Abundance with Non-Gaussian Primordial Perturbations}}, \href{https://doi.org/10.1088/1475-7516/2009/01/010}{\emph{JCAP} {\bfseries 01} (2009) 010} [\href{https://arxiv.org/abs/0809.0506}{{\ttfamily 0809.0506}}].

\bibitem{Pisani:2019cvo}
A.~Pisani et~al., \emph{{Cosmic voids: a novel probe to shed light on our Universe}},  \href{https://arxiv.org/abs/1903.05161}{{\ttfamily 1903.05161}}.

\bibitem{Banerjee:2020umh}
A.~Banerjee and T.~Abel, \emph{{Nearest neighbour distributions: New statistical measures for cosmological clustering}}, \href{https://doi.org/10.1093/mnras/staa3604}{\emph{Mon. Not. Roy. Astron. Soc.} {\bfseries 500} (2020) 5479} [\href{https://arxiv.org/abs/2007.13342}{{\ttfamily 2007.13342}}].

\bibitem{Coulton:2023ouk}
W.R.~Coulton, T.~Abel and A.~Banerjee, \emph{{Small-scale signatures of primordial non-Gaussianity in k-nearest neighbour cumulative distribution functions}}, \href{https://doi.org/10.1093/mnras/stae2108}{\emph{Mon. Not. Roy. Astron. Soc.} {\bfseries 534} (2024) 1621} [\href{https://arxiv.org/abs/2309.15151}{{\ttfamily 2309.15151}}].

\bibitem{Lippich:2020vpy}
M.~Lippich and A.G.~S\'anchez, \emph{{medusa: Minkowski functionals estimated from Delaunay tessellations of the three-dimensional large-scale structure}}, \href{https://doi.org/10.1093/mnras/stab2820}{\emph{Mon. Not. Roy. Astron. Soc.} {\bfseries 508} (2021) 3771} [\href{https://arxiv.org/abs/2012.08529}{{\ttfamily 2012.08529}}].

\bibitem{Liu:2023qrj}
W.~Liu, A.~Jiang and W.~Fang, \emph{{Probing massive neutrinos with the Minkowski functionals of the galaxy distribution}}, \href{https://doi.org/10.1088/1475-7516/2023/09/037}{\emph{JCAP} {\bfseries 09} (2023) 037} [\href{https://arxiv.org/abs/2302.08162}{{\ttfamily 2302.08162}}].

\bibitem{Jiang:2023nzz}
A.~Jiang, W.~Liu, W.~Fang, B.~Li, C.~Barrera-Hinojosa and Y.~Zhang, \emph{{Minkowski functionals of large-scale structure as a probe of modified gravity}}, \href{https://doi.org/10.1103/PhysRevD.109.083537}{\emph{Phys. Rev. D} {\bfseries 109} (2024) 083537} [\href{https://arxiv.org/abs/2305.04520}{{\ttfamily 2305.04520}}].

\bibitem{SimBIG:2023ywd}
{\scshape SimBIG} collaboration, \emph{{Field-level simulation-based inference of galaxy clustering with convolutional neural networks}}, \href{https://doi.org/10.1103/PhysRevD.109.083536}{\emph{Phys. Rev. D} {\bfseries 109} (2024) 083536} [\href{https://arxiv.org/abs/2310.15256}{{\ttfamily 2310.15256}}].

\bibitem{Shao:2022mzk}
H.~Shao et~al., \emph{{Robust Field-level Inference of Cosmological Parameters with Dark Matter Halos}}, \href{https://doi.org/10.3847/1538-4357/acac7a}{\emph{Astrophys. J.} {\bfseries 944} (2023) 27} [\href{https://arxiv.org/abs/2209.06843}{{\ttfamily 2209.06843}}].

\bibitem{deSanti:2023zzn}
N.S.M.~de~Santi et~al., \emph{{Robust Field-level Likelihood-free Inference with Galaxies}}, \href{https://doi.org/10.3847/1538-4357/acd1e2}{\emph{Astrophys. J.} {\bfseries 952} (2023) 69} [\href{https://arxiv.org/abs/2302.14101}{{\ttfamily 2302.14101}}].

\bibitem{Anagnostidis:2022rbs}
S.~Anagnostidis, A.~Thomsen, T.~Kacprzak, T.~Tr\"oster, L.~Biggio, A.~Refregier et~al., \emph{{Cosmology from Galaxy Redshift Surveys with PointNet}},  \href{https://arxiv.org/abs/2211.12346}{{\ttfamily 2211.12346}}.

\bibitem{Chatterjee:2024eur}
A.~Chatterjee and F.~Villaescusa-Navarro, \emph{{Cosmology from point clouds}},  \href{https://arxiv.org/abs/2405.13119}{{\ttfamily 2405.13119}}.

\bibitem{Barreira:2021ukk}
A.~Barreira, T.~Lazeyras and F.~Schmidt, \emph{{Galaxy bias from forward models: linear and second-order bias of IllustrisTNG galaxies}}, \href{https://doi.org/10.1088/1475-7516/2021/08/029}{\emph{JCAP} {\bfseries 08} (2021) 029} [\href{https://arxiv.org/abs/2105.02876}{{\ttfamily 2105.02876}}].

\bibitem{Nguyen:2024yth}
N.-M.~Nguyen, F.~Schmidt, B.~Tucci, M.~Reinecke and A.~Kosti\'c, \emph{{How Much Information Can Be Extracted from Galaxy Clustering at the Field Level?}}, \href{https://doi.org/10.1103/PhysRevLett.133.221006}{\emph{Phys. Rev. Lett.} {\bfseries 133} (2024) 221006} [\href{https://arxiv.org/abs/2403.03220}{{\ttfamily 2403.03220}}].

\bibitem{Babic:2024wph}
I.~Babi\'c, F.~Schmidt and B.~Tucci, \emph{{Straightening the Ruler: Field-Level Inference of the BAO Scale with LEFTfield}},  \href{https://arxiv.org/abs/2407.01524}{{\ttfamily 2407.01524}}.

\bibitem{Kostic:2022vok}
A.~Kosti\'c, N.-M.~Nguyen, F.~Schmidt and M.~Reinecke, \emph{{Consistency tests of field level inference with the EFT likelihood}}, \href{https://doi.org/10.1088/1475-7516/2023/07/063}{\emph{JCAP} {\bfseries 07} (2023) 063} [\href{https://arxiv.org/abs/2212.07875}{{\ttfamily 2212.07875}}].

\bibitem{Wilding:2020oza}
G.~Wilding, K.~Nevenzeel, R.~van~de Weygaert, G.~Vegter, P.~Pranav, B.J.T.~Jones et~al., \emph{{Persistent homology of the cosmic web \textendash{} I. Hierarchical topology in \ensuremath{\Lambda}CDM cosmologies}}, \href{https://doi.org/10.1093/mnras/stab2326}{\emph{Mon. Not. Roy. Astron. Soc.} {\bfseries 507} (2021) 2968} [\href{https://arxiv.org/abs/2011.12851}{{\ttfamily 2011.12851}}].

\bibitem{2010MNRAS.408.2163A}
M.A.~{Arag{\'o}n-Calvo}, R.~{van de Weygaert} and B.J.T.~{Jones}, \emph{{Multiscale phenomenology of the cosmic web}}, \href{https://doi.org/10.1111/j.1365-2966.2010.17263.x}{\emph{mnras} {\bfseries 408} (2010) 2163} [\href{https://arxiv.org/abs/1007.0742}{{\ttfamily 1007.0742}}].

\bibitem{Sousbie2011}
T.~{Sousbie}, C.~{Pichon} and H.~{Kawahara}, \emph{{The persistent cosmic web and its filamentary structure - II. Illustrations}}, \href{https://doi.org/10.1111/j.1365-2966.2011.18395.x}{\emph{mnras} {\bfseries 414} (2011) 384} [\href{https://arxiv.org/abs/1009.4014}{{\ttfamily 1009.4014}}].

\bibitem{10.1093/mnras/stw2862}
P.~Pranav, H.~Edelsbrunner, R.~van de Weygaert, G.~Vegter, M.~Kerber, B.J.T.~Jones et~al., \emph{The topology of the cosmic web in terms of persistent betti numbers}, \href{https://doi.org/10.1093/mnras/stw2862}{\emph{Monthly Notices of the Royal Astronomical Society} {\bfseries 465} (2016) 4281}.

\bibitem{Cole:2020gzt}
A.~Cole, M.~Biagetti and G.~Shiu, \emph{{Topological Echoes of Primordial Physics in the Universe at Large Scales}},  in \emph{{34th Conference on Neural Information Processing Systems}}, 12, 2020 [\href{https://arxiv.org/abs/2012.03616}{{\ttfamily 2012.03616}}].

\bibitem{Biagetti:2022qjl}
M.~Biagetti, J.~Calles, L.~Castiblanco, A.~Cole and J.~Nore\~na, \emph{{Fisher forecasts for primordial non-Gaussianity from persistent homology}}, \href{https://doi.org/10.1088/1475-7516/2022/10/002}{\emph{JCAP} {\bfseries 10} (2022) 002} [\href{https://arxiv.org/abs/2203.08262}{{\ttfamily 2203.08262}}].

\bibitem{Yip:2024hlz}
J.H.T.~Yip, M.~Biagetti, A.~Cole, K.~Viswanathan and G.~Shiu, \emph{{Cosmology with Persistent Homology: a Fisher Forecast}},  \href{https://arxiv.org/abs/2403.13985}{{\ttfamily 2403.13985}}.

\bibitem{Biagetti:2020skr}
M.~Biagetti, A.~Cole and G.~Shiu, \emph{{The Persistence of Large Scale Structures I: Primordial non-Gaussianity}}, \href{https://doi.org/10.1088/1475-7516/2021/04/061}{\emph{JCAP} {\bfseries 04} (2021) 061} [\href{https://arxiv.org/abs/2009.04819}{{\ttfamily 2009.04819}}].

\bibitem{Feldbrugge_2019}
J.~Feldbrugge, M.~van Engelen, R.~van~de Weygaert, P.~Pranav and G.~Vegter, \emph{Stochastic homology of gaussian vs. non-gaussian random fields: graphs towards betti numbers and persistence diagrams}, \href{https://doi.org/10.1088/1475-7516/2019/09/052}{\emph{Journal of Cosmology and Astroparticle Physics} {\bfseries 2019} (2019) 052–052}.

\bibitem{Heydenreich:2020hrr}
S.~Heydenreich, B.~Br\"uck and J.~Harnois-D\'eraps, \emph{{Persistent homology in cosmic shear: constraining parameters with topological data analysis}}, \href{https://doi.org/10.1051/0004-6361/202039048}{\emph{Astron. Astrophys.} {\bfseries 648} (2021) A74} [\href{https://arxiv.org/abs/2007.13724}{{\ttfamily 2007.13724}}].

\bibitem{Heydenreich:2022dci}
S.~Heydenreich, B.~Br\"uck, P.~Burger, J.~Harnois-D\'eraps, S.~Unruh, T.~Castro et~al., \emph{{Persistent homology in cosmic shear - II. A tomographic analysis of DES-Y1}}, \href{https://doi.org/10.1051/0004-6361/202243868}{\emph{Astron. Astrophys.} {\bfseries 667} (2022) A125} [\href{https://arxiv.org/abs/2204.11831}{{\ttfamily 2204.11831}}].

\bibitem{Kanafi:2023twi}
M.H.J.~Kanafi, S.~Ansarifard and S.M.S.~Movahed, \emph{{Imprint of massive neutrinos on Persistent Homology of large-scale structure}},  \href{https://arxiv.org/abs/2311.13520}{{\ttfamily 2311.13520}}.

\bibitem{books/daglib/0025666}
H.~Edelsbrunner and J.~Harer, \emph{Computational Topology - an Introduction.}, American Mathematical Society (2010).

\bibitem{carlsson2021topological}
G.~Carlsson and M.~Vejdemo-Johansson, \emph{Topological Data Analysis with Applications}, Topological Data Analysis with Applications, Cambridge University Press (2021).

\bibitem{wasserman2016topologicaldataanalysis}
L.~Wasserman, \emph{Topological data analysis},  2016.

\bibitem{10.1145/174462.156635}
H.~Edelsbrunner and E.P.~M\"{u}cke, \emph{Three-dimensional alpha shapes}, \href{https://doi.org/10.1145/174462.156635}{\emph{ACM Trans. Graph.} {\bfseries 13} (1994) 43–72}.

\bibitem{1056714}
H.~Edelsbrunner, D.~Kirkpatrick and R.~Seidel, \emph{On the shape of a set of points in the plane}, \href{https://doi.org/10.1109/TIT.1983.1056714}{\emph{IEEE Transactions on Information Theory} {\bfseries 29} (1983) 551}.

\bibitem{gudhi:urm}
T.G.~Project, \emph{GUDHI User and Reference Manual}, GUDHI Editorial Board, 3.11.0~ed. (2025).

\bibitem{cgal:eb-19a}
{The CGAL Project}, \emph{{CGAL} User and Reference Manual}, {CGAL Editorial Board}, {4.14}~ed. (2019).

\bibitem{vandeweygaert2013alphabettimegaparsecuniverse}
R.~van~de Weygaert, G.~Vegter, H.~Edelsbrunner, B.J.T.~Jones, P.~Pranav, C.~Park et~al., \emph{Alpha, betti and the megaparsec universe: on the topology of the cosmic web},  2013.

\bibitem{bermejo2024topologicalbiashaloestrace}
R.~Bermejo, G.~Wilding, R.~van~de Weygaert, B.J.T.~Jones, G.~Vegter and K.~Efstathiou, \emph{Topological bias: How haloes trace structural patterns in the cosmic web},  2024.

\bibitem{Villaescusa-Navarro:2019bje}
F.~Villaescusa-Navarro et~al., \emph{{The Quijote simulations}}, \href{https://doi.org/10.3847/1538-4365/ab9d82}{\emph{Astrophys. J. Suppl.} {\bfseries 250} (2020) 2} [\href{https://arxiv.org/abs/1909.05273}{{\ttfamily 1909.05273}}].

\bibitem{Coulton:2022qbc}
W.R.~Coulton, F.~Villaescusa-Navarro, D.~Jamieson, M.~Baldi, G.~Jung, D.~Karagiannis et~al., \emph{{Quijote-PNG: Simulations of Primordial Non-Gaussianity and the Information Content of the Matter Field Power Spectrum and Bispectrum}}, \href{https://doi.org/10.3847/1538-4357/aca8a7}{\emph{Astrophys. J.} {\bfseries 943} (2023) 64} [\href{https://arxiv.org/abs/2206.01619}{{\ttfamily 2206.01619}}].

\bibitem{Jung:2023kjh}
G.~Jung et~al., \emph{{Quijote-PNG: The Information Content of the Halo Mass Function}}, \href{https://doi.org/10.3847/1538-4357/acfe70}{\emph{Astrophys. J.} {\bfseries 957} (2023) 50} [\href{https://arxiv.org/abs/2305.10597}{{\ttfamily 2305.10597}}].

\bibitem{Sefusatti:2015aex}
E.~Sefusatti, M.~Crocce, R.~Scoccimarro and H.~Couchman, \emph{{Accurate Estimators of Correlation Functions in Fourier Space}}, \href{https://doi.org/10.1093/mnras/stw1229}{\emph{Mon. Not. Roy. Astron. Soc.} {\bfseries 460} (2016) 3624} [\href{https://arxiv.org/abs/1512.07295}{{\ttfamily 1512.07295}}].

\bibitem{Jeffrey:2020itg}
N.~Jeffrey and B.D.~Wandelt, \emph{{Solving high-dimensional parameter inference: marginal posterior densities \& Moment Networks}},  in \emph{{34th Conference on Neural Information Processing Systems}}, 11, 2020 [\href{https://arxiv.org/abs/2011.05991}{{\ttfamily 2011.05991}}].

\bibitem{Villaescusa-Navarro:2021pkb}
F.~Villaescusa-Navarro et~al., \emph{{Multifield Cosmology with Artificial Intelligence}},  \href{https://arxiv.org/abs/2109.09747}{{\ttfamily 2109.09747}}.

\bibitem{Villaescusa-Navarro:2021cni}
F.~Villaescusa-Navarro et~al., \emph{{Robust marginalization of baryonic effects for cosmological inference at the field level}},  \href{https://arxiv.org/abs/2109.10360}{{\ttfamily 2109.10360}}.

\bibitem{Villanueva-Domingo:2021dun}
P.~Villanueva-Domingo, F.~Villaescusa-Navarro, D.~Angl\'es-Alc\'azar, S.~Genel, F.~Marinacci, D.N.~Spergel et~al., \emph{{Inferring Halo Masses with Graph Neural Networks}}, \href{https://doi.org/10.3847/1538-4357/ac7aa3}{\emph{Astrophys. J.} {\bfseries 935} (2022) 30} [\href{https://arxiv.org/abs/2111.08683}{{\ttfamily 2111.08683}}].

\bibitem{Wang:2022zpv}
B.Y.~Wang, A.~Pisani, F.~Villaescusa-Navarro and B.D.~Wandelt, \emph{{Machine-learning Cosmology from Void Properties}}, \href{https://doi.org/10.3847/1538-4357/aceaf6}{\emph{Astrophys. J.} {\bfseries 955} (2023) 131} [\href{https://arxiv.org/abs/2212.06860}{{\ttfamily 2212.06860}}].

\bibitem{Villanueva-Domingo:2022rvn}
P.~Villanueva-Domingo and F.~Villaescusa-Navarro, \emph{{Learning Cosmology and Clustering with Cosmic Graphs}}, \href{https://doi.org/10.3847/1538-4357/ac8930}{\emph{Astrophys. J.} {\bfseries 937} (2022) 115} [\href{https://arxiv.org/abs/2204.13713}{{\ttfamily 2204.13713}}].

\bibitem{Perez:2022nlv}
L.A.~Perez, S.~Genel, F.~Villaescusa-Navarro, R.S.~Somerville, A.~Gabrielpillai, D.~Angl\'es-Alc\'azar et~al., \emph{{Constraining Cosmology with Machine Learning and Galaxy Clustering: The CAMELS-SAM Suite}}, \href{https://doi.org/10.3847/1538-4357/accd52}{\emph{Astrophys. J.} {\bfseries 954} (2023) 11} [\href{https://arxiv.org/abs/2204.02408}{{\ttfamily 2204.02408}}].

\bibitem{Villaescusa-Navarro:2022twv}
F.~Villaescusa-Navarro et~al., \emph{{Cosmology with One Galaxy?}}, \href{https://doi.org/10.3847/1538-4357/ac5d3f}{\emph{Astrophys. J.} {\bfseries 929} (2022) 132} [\href{https://arxiv.org/abs/2201.02202}{{\ttfamily 2201.02202}}].

\bibitem{Chawak:2023bil}
C.~Chawak, F.~Villaescusa-Navarro, N.E.~Rojas, Y.~Ni, C.~Hahn and D.~Angles-Alcazar, \emph{{Cosmology with Multiple Galaxies}}, \href{https://doi.org/10.3847/1538-4357/ad4969}{\emph{Astrophys. J.} {\bfseries 969} (2024) 105} [\href{https://arxiv.org/abs/2309.12048}{{\ttfamily 2309.12048}}].

\bibitem{deSanti:2023rsw}
N.S.M.~de~Santi et~al., \emph{{Field-level simulation-based inference with galaxy catalogs: the impact of systematic effects}},  \href{https://arxiv.org/abs/2310.15234}{{\ttfamily 2310.15234}}.

\bibitem{Gondhalekar:2024iqm}
Y.~Gondhalekar and K.~Moriwaki, \emph{{Convolutional Vision Transformer for Cosmology Parameter Inference}},  11, 2024 [\href{https://arxiv.org/abs/2411.14392}{{\ttfamily 2411.14392}}].

\bibitem{grinsztajn2022tree}
L.~Grinsztajn, E.~Oyallon and G.~Varoquaux, \emph{Why do tree-based models still outperform deep learning on typical tabular data?}, {\emph{Advances in neural information processing systems} {\bfseries 35} (2022) 507}.

\bibitem{Lazanu:2021tdl}
A.~Lazanu, \emph{{Extracting cosmological parameters from N-body simulations using machine learning techniques}}, \href{https://doi.org/10.1088/1475-7516/2021/09/039}{\emph{JCAP} {\bfseries 09} (2021) 039} [\href{https://arxiv.org/abs/2106.11061}{{\ttfamily 2106.11061}}].

\bibitem{Chen_2016}
T.~Chen and C.~Guestrin, \emph{Xgboost: A scalable tree boosting system},  in \emph{Proceedings of the 22nd ACM SIGKDD International Conference on Knowledge Discovery and Data Mining}, KDD ’16, p.~785–794, ACM, Aug., 2016, \href{https://doi.org/10.1145/2939672.2939785}{DOI}.

\bibitem{Babich:2004gb}
D.~Babich, P.~Creminelli and M.~Zaldarriaga, \emph{{The Shape of non-Gaussianities}}, \href{https://doi.org/10.1088/1475-7516/2004/08/009}{\emph{JCAP} {\bfseries 08} (2004) 009} [\href{https://arxiv.org/abs/astro-ph/0405356}{{\ttfamily astro-ph/0405356}}].

\bibitem{Valogiannis:2021chp}
G.~Valogiannis and C.~Dvorkin, \emph{{Towards an optimal estimation of cosmological parameters with the wavelet scattering transform}}, \href{https://doi.org/10.1103/PhysRevD.105.103534}{\emph{Phys. Rev. D} {\bfseries 105} (2022) 103534} [\href{https://arxiv.org/abs/2108.07821}{{\ttfamily 2108.07821}}].

\bibitem{Massara:2024cvu}
E.~Massara, C.~Hahn, M.~Eickenberg, S.~Ho, J.~Hou, P.~Lemos et~al., \emph{{{\textbackslash{}sc SimBIG}: Cosmological Constraints using Simulation-Based Inference of Galaxy Clustering with Marked Power Spectra}},  \href{https://arxiv.org/abs/2404.04228}{{\ttfamily 2404.04228}}.

\bibitem{Makinen:2022jsc}
T.L.~Makinen, T.~Charnock, P.~Lemos, N.~Porqueres, A.~Heavens and B.D.~Wandelt, \emph{{The Cosmic Graph: Optimal Information Extraction from Large-Scale Structure using Catalogues}},  \href{https://arxiv.org/abs/2207.05202}{{\ttfamily 2207.05202}}.

\bibitem{Hortua:2023kuw}
H.J.~Hort\'ua, L.A.~Garc\'\i{}a and L.~Casta\~neda C., \emph{{Constraining cosmological parameters from N-body simulations with variational Bayesian neural networks}}, \href{https://doi.org/10.3389/fspas.2023.1139120}{\emph{Front. Astron. Space Sci.} {\bfseries 10} (2023) 1139120} [\href{https://arxiv.org/abs/2301.03991}{{\ttfamily 2301.03991}}].

\bibitem{Ho:2024whi}
M.~Ho et~al., \emph{{LtU-ILI: An All-in-One Framework for Implicit Inference in Astrophysics and Cosmology}}, \href{https://doi.org/10.33232/001c.120559}{\emph{Open J. Astrophys.} {\bfseries 7} (2024) 001c.120559} [\href{https://arxiv.org/abs/2402.05137}{{\ttfamily 2402.05137}}].

\bibitem{Cuesta-Lazaro:2023zuk}
C.~Cuesta-Lazaro and S.~Mishra-Sharma, \emph{{Point cloud approach to generative modeling for galaxy surveys at the field level}}, \href{https://doi.org/10.1103/PhysRevD.109.123531}{\emph{Phys. Rev. D} {\bfseries 109} (2024) 123531} [\href{https://arxiv.org/abs/2311.17141}{{\ttfamily 2311.17141}}].

\bibitem{Min:2024dgd}
Z.~Min et~al., \emph{{Deep learning for cosmological parameter inference from a dark matter halo density field}}, \href{https://doi.org/10.1103/PhysRevD.110.063531}{\emph{Phys. Rev. D} {\bfseries 110} (2024) 063531} [\href{https://arxiv.org/abs/2404.09483}{{\ttfamily 2404.09483}}].

\bibitem{zaheer2017deep}
M.~Zaheer, S.~Kottur, S.~Ravanbakhsh, B.~Poczos, R.R.~Salakhutdinov and A.J.~Smola, \emph{Deep sets}, {\emph{Advances in neural information processing systems} {\bfseries 30} (2017) }.

\bibitem{carriere2020perslay}
M.~Carri{\`e}re, F.~Chazal, Y.~Ike, T.~Lacombe, M.~Royer and Y.~Umeda, \emph{Perslay: A neural network layer for persistence diagrams and new graph topological signatures},  in \emph{International Conference on Artificial Intelligence and Statistics}, pp.~2786--2796, PMLR, 2020.

\bibitem{cranmer2020frontier}
K.~Cranmer, J.~Brehmer and G.~Louppe, \emph{The frontier of simulation-based inference}, {\emph{Proceedings of the National Academy of Sciences} {\bfseries 117} (2020) 30055}.

\bibitem{Reza:2024djq}
M.~Reza, Y.~Zhang, C.~Avestruz, L.E.~Strigari, S.~Shevchuk and F.~Villaescusa-Navarro, \emph{{Constraining Cosmology with Simulation-based inference and Optical Galaxy Cluster Abundance}},  \href{https://arxiv.org/abs/2409.20507}{{\ttfamily 2409.20507}}.

\bibitem{Miller:2021hys}
B.K.~Miller, A.~Cole, P.~Forr\'e, G.~Louppe and C.~Weniger, \emph{{Truncated Marginal Neural Ratio Estimation}},  in \emph{{35th Conference on Neural Information Processing Systems}}, 7, 2021, \href{https://doi.org/10.5281/zenodo.5043706}{DOI} [\href{https://arxiv.org/abs/2107.01214}{{\ttfamily 2107.01214}}].

\bibitem{Mootoovaloo:2024sao}
A.~Mootoovaloo, C.~Garc\'\i{}a-Garc\'\i{}a, D.~Alonso and J.~Ruiz-Zapatero, \emph{{$\mathtt{emuflow}$: Normalising Flows for Joint Cosmological Analysis}},  \href{https://arxiv.org/abs/2409.01407}{{\ttfamily 2409.01407}}.

\bibitem{Makinen:2024xph}
T.L.~Makinen, C.~Sui, B.D.~Wandelt, N.~Porqueres and A.~Heavens, \emph{{Hybrid Summary Statistics}},  \href{https://arxiv.org/abs/2410.07548}{{\ttfamily 2410.07548}}.

\bibitem{khullar2022digs}
G.~Khullar, B.~Nord, A.~{\'C}iprijanovi{\'c}, J.~Poh and F.~Xu, \emph{Digs: deep inference of galaxy spectra with neural posterior estimation}, {\emph{Machine Learning: Science and Technology} {\bfseries 3} (2022) 04LT04}.

\bibitem{Lehman:2024vyl}
K.~Lehman, S.~Krippendorf, J.~Weller and K.~Dolag, \emph{{Learning Optimal and Interpretable Summary Statistics of Galaxy Catalogs with SBI}},  \href{https://arxiv.org/abs/2411.08957}{{\ttfamily 2411.08957}}.

\bibitem{Saxena:2024rhu}
A.~Saxena, P.D.~Meerburg, C.~Weniger, E.d.L.~Acedo and W.~Handley, \emph{{Simulation-based inference of the sky-averaged 21-cm signal from CD-EoR with REACH}}, \href{https://doi.org/10.1093/rasti/rzae047}{\emph{RAS Tech. Instrum.} {\bfseries 3} (2024) 724} [\href{https://arxiv.org/abs/2403.14618}{{\ttfamily 2403.14618}}].

\bibitem{Hahn:2022zxa}
C.~Hahn, M.~Eickenberg, S.~Ho, J.~Hou, P.~Lemos, E.~Massara et~al., \emph{{A forward modeling approach to analyzing galaxy clustering with S\ensuremath{<}span class=}}, \href{https://doi.org/10.1073/pnas.2218810120}{\emph{Proc. Nat. Acad. Sci.} {\bfseries 120} (2023) e2218810120} [\href{https://arxiv.org/abs/2211.00723}{{\ttfamily 2211.00723}}].

\bibitem{Tucci:2023bag}
B.~Tucci and F.~Schmidt, \emph{{EFTofLSS meets simulation-based inference: \ensuremath{\sigma} $_{8}$ from biased tracers}}, \href{https://doi.org/10.1088/1475-7516/2024/05/063}{\emph{JCAP} {\bfseries 05} (2024) 063} [\href{https://arxiv.org/abs/2310.03741}{{\ttfamily 2310.03741}}].

\bibitem{https://doi.org/10.21231/gnt1-hw21}
{Center for High Throughput Computing}, \emph{Center for high throughput computing},  2006.
\newblock 10.21231/GNT1-HW21.

\end{thebibliography}\endgroup

\end{document}